%% file: mulchaey.tex
\documentclass[preprint]{aastex}

\slugcomment{To appear in The Astrophysical Journal Supplements}

\shorttitle{X-ray Atlas of Groups}
\shortauthors{Mulchaey et al.}

\begin{document}

\title{An X-ray Atlas of Groups of Galaxies}

\author{John S. Mulchaey}
\affil{The Observatories of The Carnegie Institution of Washington, 813 Santa
Barbara St., Pasadena, California 91101}
\email{mulchaey@ociw.edu}

\author{David S. Davis}
\affil{Joint Center for Astrophysics, University of Maryland at Baltimore County and Laboratory for High Energy Astrophysics/ Goddard Space Flight Center, Code 661.0, Greenbelt, MD 20771
}
\email{ddavis@milkyway.gsfc.nasa.gov}

\author{Richard F. Mushotzky}
\affil{Laboratory for High Energy Astrophysics, NASA/GSFC, Code 662, Greenbelt, MD 20771}
\email{mushotzky@lheavx.dnet.nasa.gov}

\and

\author{David Burstein}
\affil{Department of Physics and Astronomy, Box 871054, Arizona State University, Tempe, Arizona 85287-1504}
\email{burstein@samuri.la.asu.edu}

\begin{abstract}
A search was conducted for a hot intragroup medium in 109 low-redshift
galaxy groups observed with the ROSAT PSPC. Evidence for diffuse,
extended X-ray emission is found in at least 61 groups. Approximately
one-third of these detections have not been previously reported in the
literature. 
 Most of the groups are detected out to less than half of
the virial radius with ROSAT.  Although some spiral-rich groups do
contain an intragroup medium, diffuse emission is restricted
to groups that contain at least one early-type galaxy. 
\end{abstract}

\section{Introduction}
Most galaxies in the local universe, including our own Milky Way, are
members of poor galaxy groups \citep{GH83,T87,NW87}.  The realization
that many groups are X-ray sources has led to considerable interest in
these systems over the last decade
\citep{M93,PB93,D94,H95,PBE95,Doe95,SC95,M96a,P96,MZ98,HP00a}.  The
X-ray emission in groups is typically extended on scales of hundreds
of kiloparsecs.  X-ray spectroscopy suggests the emission mechanism is
most likely a combination of thermal bremsstrahlung and line
emission. This interpretation requires that the entire volume of
groups be filled with a hot, low-density gas often referred to as the
intragroup medium in analogy to the intracluster medium found in
richer systems.

The existence of an intragroup medium is important for many reasons.
The presence of a hot gas halo indicates that many groups are likely
real, physical systems and not simply chance superpositions or
large-scale filaments viewed edge-on. Assuming the intragroup gas is
in hydrostatic equilibrium, the total mass of the group can be
estimated \citep{M93,PB93,PBE95,M96a}.  Based on ROSAT data, the
typical group mass is approximately one-tenth the mass of a cluster
like Virgo.  However, because the number density of X-ray detected
groups is considerably higher than the number density of clusters like
Virgo, the contribution of X-ray detected groups to the total mass
density of the universe is comparable to or greater than that of rich
clusters.  A comparison of the mass in galaxies and intragroup gas to
the total group mass indicates that the known baryonic components
typically account for only 10--20\% of the total mass
\citep{M93,PB93,H95,PBE95,M96a}.  This implies that groups are
dominated by dark matter.  Extrapolating the X-ray surface brightness
profiles to the virial radius suggests that most of the baryonic mass
in groups is in the intragroup medium.  In fact, the intragroup medium
may be the dominant baryonic component in the local universe
\citep{F98}.

The intragroup medium may also hold important clues into the formation
and evolution of large-scale structure.  Numerical simulations
indicate that in the absence of non-gravitational heating, the density
profiles of groups and clusters should be nearly identical
\citep{N97}.  In this simple scenario, the relationships between X-ray
luminosity (L$_{\rm X}$), X-ray temperature (T) and optical velocity
dispersion ($\sigma$) are expected to be similar for groups and
clusters.  There is now considerable evidence for departures from such
uniformity \citep{P99,LD00,Ma00,HP00a}.  The interpretation of these
departures is currently a topic of much debate.  
One possibility is that preheating of the intragroup medium
by an early generation of stars can explain the scaling relationships
\citep{CMT98,P99,PBG00,FDP00,BEM01,B01}.  Alternative explanations include 
heating by active galaxies \citep{VS99,FT00,YM01}, internal heating
mechanisms \citep{L00} and the possibility that galaxy
formation is more efficient in groups than in clusters \citep{bryan00}.

Most of our current understanding of the intragroup medium is based on
observations of groups with ROSAT. The low internal background, large
field of view, and good sensitivity to soft X-rays made the ROSAT PSPC
detectors ideal instruments for studies of nearby
groups. Unfortunately, large systematic surveys of well-selected
groups were not performed during the limited lifetime of the PSPC
detectors. However, a large number of groups were observed
serendipitously. Although ROSAT results have been reported for many of
these groups, a significant fraction of the groups in the ROSAT
database have not been previously published.  Furthermore, the derived
X-ray properties of groups are somewhat sensitive to assumptions made
during the data reduction and analysis. For this reason, different
authors often derive vastly different results from the same ROSAT
dataset. Therefore, constructing large samples from the literature can
be problematic.  In an attempt to produce a more uniform sample, we
have reanalyzed the ROSAT PSPC data for all poor groups previously
reported in the literature along with a large number of nearby groups
that have not previously been published.

We describe the sample selection in \S 2 and the data reduction 
in \S 3. For those groups with detected diffuse emission, we have
extracted an X-ray spectrum (\S 4). Fits to the surface
brightness profiles of the groups are described in \S 5.
In \S 6 we examine the morphological composition of 
the X-ray detected groups. Finally, a summary is provided in \S 7. 

\section{Sample}
The ROSAT mission consisted of two main scientific phases, a
six-month, all-sky survey with a mean exposure time of approximately
400 seconds and longer pointings of individual targets (the so-called
`pointed mode').  While the all-sky survey data are deep enough to
detect the most X-ray luminous groups \citep{E94,H95,Ma97,Ma00}, it is
not possible to carry-out detailed spatial and spectral studies with
this dataset. As the `pointed mode' observations allow more detailed
studies, we restrict our analysis to groups observed during this phase
of the ROSAT mission.

We select our groups from a number of different sources. We start by
including all groups that were in previous ROSAT `pointed-mode'
studies.  Many of the groups come from the compact group catalog of
\citet{H82} or from the loose group studies of \citet{M96a},
\citet{MZ98} and \citet{HP00a}.  Additional targets were found by
cross-correlating the ROSAT observation log with the positions of
optically-selected groups in the catalogs of \citet{HG82},
\citet{GH83}, \citet{M89}, \citet{N93} and \citet{G93}.  These
catalogs also include richer galaxy systems (i.e. clusters).  For the
purpose of this paper, we only include \lq\lq groups\rq\rq \ with
velocity dispersions less than 600 km/s or an intragroup medium
temperature less than 2 keV. A few of the systems we have encountered
satisfy one, but not the other criterion. We have opted to include
these systems in our sample.  In general, we only include groups whose
centers fall within 20$'$ of the center of the ROSAT field. The only
exceptions are a few Hickson Compact Groups, which were previously
included in \citet{P96} and are included here for comparison. We have
not included HCG 73 in our sample, as it is right at the edge of the
ROSAT field. We also eliminated any groups that are projected in front
of or behind rich clusters.

The final sample of 109 groups is given in Table 1. The name of each
group is given in the first column, followed by the Right Ascension
and Declination (columns 2 and 3), the number of cataloged galaxies
(column 4), redshift (column 5) and velocity dispersion (column 6).
For groups with extended X-ray emission, the right ascensions and
declinations are generally given for the peak of the X-ray
emission, which was determined in each case 
from the ROSAT image smoothed with a Gaussian profile of width 30$''$.
 For a few X-ray detected groups, no clear flux peak
exists. In these cases, the coordinates refer to the approximate
center of the emission morphology.  For the non-detected groups, we
adopt the coordinates given in the optical group catalogs.  Column 7
gives the spiral fraction, defined as the fraction of cataloged
members that are late-type galaxies (i.e. spirals and irregulars). We
note that the quantities listed in Table 1 have not been calculated to a
single minimum absolute magnitude.  For most of the groups, the number
of members, velocity dispersion and
spiral fractions are based on the few brightest ($\sim$ L$_{\star}$)
galaxies in the original redshift surveys. However, a few groups have
morphologies and membership data down to significantly lower
luminosities \citep{ZM98,Ma99}. Therefore, some caution should be used when 
comparing the optical properties of individual groups. 
Column 8 gives the references for the quantities in
columns 2-7. The last column gives the cross references for the group
in optical group catalogs.


\section{Data Reduction and Analysis}
To search for a hot intragroup medium, we follow
the reduction method outlined in \citet{MZ98}. First, to remove times
of high particle background, we discard any data taken when the master
veto rate is greater than 170 counts s$^{-1}$. For each ROSAT
observation, six images are created corresponding to the R2-R7 energy
bands given in \citet{S94}. Each of these images is corrected for
vignetting using energy-dependent exposure maps. These individual
images are combined to produce two final images; one corresponding to
the \lq\lq standard\rq\rq \ energy range $\sim$ 0.45--2.05 keV used in
our previous analysis \citep{M96a,MZ98} and one that includes the
\lq\lq softer\rq\rq \ energy bands ($\sim$ 0.15-2.05 keV).  We created
the softer energy band images with the hope of detecting very cool
intragroup gas not visible in the standard images.  However, in all
cases where emission was found in the softer images, it was also
detected in the standard image. Given that the background level is
much higher in the softer energy bands, all the analysis presented in
this paper is based on the standard images.

Before searching for diffuse intragroup gas, we remove emission
associated with point sources. Point sources are identified using the
task \lq\lq DETECT\rq\rq \ in the Extended Object and X-ray Background
Analysis software \citep{S94}.  We note that in the present context,
\lq\lq point source\rq\rq \ refers to a source that appears point-like
at the resolution of the ROSAT PSPC (FWHM $\sim$ 30$''$ on-axis and
considerably worse off-axis).  Emission from point sources is removed
by excluding a circular region around each source with a radius 1.5
times the radius that encircles 90\% of the source flux. This
exclusion radius corresponds to 1.5$'$ for sources on-axis
\citep{has92}. In a few cases, it was necessary to use a slightly 
larger aperture to remove the flux of more extended sources in the field
(such as a very nearby galaxy or a background cluster of galaxies).
In most cases where diffuse emission is detected, the
emission is centered on a luminous elliptical or S0 galaxy.  While the
\lq\lq DETECT\rq\rq \ software will usually identify the peak of the
extended emission as a \lq\lq point source\rq\rq \, we do not exclude
the emission in these cases.

A small fraction of the groups were observed by ROSAT in multiple
pointings. In these cases, we reduced the data for each pointing as
described above and then combined the images before running the task
\lq\lq DETECT\rq\rq \ .

Once the point source emission is removed, the unsmoothed images are
examined to determine if a diffuse emission component is present.  In
addition to a visual inspection, we also create an
azimuthally-averaged surface brightness profile for each group to
determine if there is excess emission above the background
level. Finally, we estimate the excess group counts by adding up the
total counts within a radius of 200 h$_{\rm 100}^{-1}$ kpc of the
group center and subtracting the vignetting-corrected background
counts estimated from an outer region of the field. In all cases where
the visual inspection implied the presence of diffuse emission, both
the surface brightness profile and excess count rate are consistent
with the existence of intragroup gas. Furthermore, in all cases where
the surface brightness profile suggested the presence of diffuse gas,
the visual inspection also indicated a diffuse component.

The results of our diffuse emission search and the spectral properties of the 
diffuse gas are summarized for each group in
Table 2. The second column in Table 2 gives the total useful ROSAT exposure
time in seconds. 
The third column indicates whether diffuse emission was detected
or not.
Groups that clearly show evidence for diffuse emission are
labeled \lq\lq DE\rq\rq \ (for diffuse emission), while
the non-detected groups are labeled \lq\lq UL\rq\rq \ (for upper
limit).  A few of the groups that contain a diffuse component also
contain a significant contribution from a central AGN. These hybrid
objects are labeled \lq\lq DE/AGN\rq\rq \ in Table 2. For a few
groups, emission is detected, but the nature of the emission is such
that it might not be associated with the group or might not be truly
diffuse.  These cases are labeled \lq\lq DE?\rq\rq \ in column 3.  In
HCG 33, a very faint diffuse component is detected, but it is offset
from the optical center of the group by nearly 18.5$'$ (approximately
400 kpc). Thus, it is unclear whether this emission is associated with
the group or a foreground/background object.  There are several 
patches of diffuse
emission in the field of HCG 22, but the brightest emission 
is centered on the background
galaxy NGC 1192 and therefore most likely not associated with HCG 22.
Several other groups (HCG 15, HCG 35, HCG 48, HCG 57) are clearly
detected in the ROSAT image, but are so far off-axis that it is not
possible to cleanly separate an intragroup gas component from 
emission associated with individual galaxies. Column 4 in Table 2 lists the 
Galactic neutral column density towards the line of site as given 
in \cite{DL90}, while column 5
gives the maximum radius of X-ray detection determined from
examining the surface brightness profile of each group.
Columns 6 and 7 give the temperature
and metal abundance, respectively,
 derived from fits to the ROSAT spectra (see \S4).
The distance to each group (column 8) has been calculated from the recessional
velocity with a correction for infall to Virgo and the Great Attractor
\citep{F96}. Finally, the bolometric luminosity is given in the last
column in Table 2.

In Figure 1, we overlay contours of the X-ray emission on the STScI
Digitized Sky Survey images for all groups that show evidence for
diffuse emission. As has been noted in previous studies
\citep{M96a,MZ98,HP00a}, the diffuse X-ray emission in groups is
almost always peaked on an elliptical or S0 galaxy. The emission
associated with the central galaxy has been included in Figure 1
except for the cases where an AGN component clearly dominates.  All
other emission due to point sources in the field has been removed.

Most of the groups in our survey have round, symmetrical
morphologies. These morphologies are consistent with the groups being
virialized. However, some groups have very irregular, clumpy
morphologies that are not centered on any particular galaxy (e.g. HCG
16, HCG 37, SHK 202, NGC 7777). These groups are likely not virialized
\citep{ZM98,DM99}.  Five groups have clearly defined bi-modal
X-ray distributions (NGC 507, NGC 1407, NGC 3607, NGC 4065 and NGC
7619). In all these cases, the second X-ray peak is also centered on a
luminous early-type galaxy.  These objects could be groups in the
process of merging.  Several of the high redshift X-ray groups
discovered in the deep Chandra pointing of the Hubble Deep Field North
also show evidence for bi-modal X-ray morphologies
\citep{B02}. Although the number of high redshift groups is too small
to make any concrete statements, the HDF North data suggests that
bi-modal groups may be more common at high redshift (bi-modal groups
make up less than 10\% of our X-ray detected groups). This idea should
be testable in the near future as many more high redshift groups are
discovered in deep Chandra and XMM-Newton pointings.

\section{Spectral Analysis}

For each group with a diffuse emission component, we have attempted to
extract a spectrum.  The spectra are extracted using a radius that
corresponds to the maximum radius of detection determined from
examining the surface brightness profile of each group (R$_{\rm X}$; see
column 5 in Table 2). Point sources
are excluded from the spectrum using the exclusion radii described
above (\S3). The vignetting-corrected background is estimated in most
cases from an annulus with inner radius 36$'$ and outer radius
42$'$. In a few cases, a different background region is used to avoid
the presence of a luminous cluster of galaxies.

To derive the spectral properties of the diffuse gas, we fit the
spectra using the software package XSPEC. For those groups with
multiple ROSAT pointings, we fit the individual spectra from each
pointing simultaneously.  We fit the spectra with an absorbed MEKAL
plasma model \citep{M85,KM93,L95}. 
We fix the absorbing
column at the Galactic value given in \citet{DL90}.  For most of the
groups, we are unable to constrain the metallicity of the gas with the
ROSAT data. In these cases, we fix the gas metallicity to 0.3 solar.
For consistency with previous X-ray studies, we adopt the old
photospheric value for the solar Fe abundance. To renormalize to the
meteoritic value, our metallicity measurements should be increased by
a factor of approximately $\sim$ 1.44.  For those groups with
sufficient data to fit the metallicity explicitly, we also perform
spectral fits with the metallicity fixed at 0.3 solar.

The quoted errors on the temperature and metallicity measurements 
are at the 90\%
confidence level.  In the UGC 1651 group, the temperature cannot be
adequately constrained and no errors are given.  The derived
temperatures range from $\sim$ 0.3 keV to over 2 keV ($\sim$ 3.5
$\times$ 10$^6$K - 2.3 $\times$ 10$^7$ K), with the vast majority of
groups centered around 1 keV ($\sim$ 1.2 $\times$ 10$^7$ K; see Figure
2).  Groups cooler than $\sim$ 0.3 keV would be difficult to detect
with ROSAT because of the high surface brightness of the X-ray
background at these soft energies and heavy absorption by Galactic HI
\citep{M96b}.  Most systems above 2 keV would normally be considered
clusters and were thus eliminated from our original sample.  To
examine how the derived temperatures depend on the assumed
metallicity, we compare the derived temperatures for the groups where
both a fixed metallicity and variable metallicity model were used
(Figure 3). In general, the temperatures are not very sensitive to the
assumed metallicity.  An exception to this is the NGC 383 group. In
this case, fixing the metallicity at a value of 0.3 solar leads to a
temperature over twice that found in the variable abundance case.

In general, there is good agreement between the temperatures we derive
and previous studies of the same groups with ROSAT. Figure 4 shows a
comparison of the temperatures from our survey with the temperatures
in \citet{HP00b}. The agreement is good despite the fact that
different apertures were used for the spectral extractions in many
cases. In particular, the \citet{HP00b} temperatures for most of the
Hickson Compact Groups were derived using a fixed physical size on the
sky, while we use the maximum radius of detection for our extractions.

There is less agreement between our derived temperatures and those
measured with the ASCA X-ray telescope.  Figure 5 compares the
temperatures derived from these two telescopes for all groups in our
survey with published ASCA measurements. While the ROSAT and ASCA
measurements are similar for cool groups, there are big differences
for hotter systems. Above 1 keV, the ASCA temperatures tend to be
higher than the ROSAT temperatures by as much as 50\%. This trend has
been previously noted by \citet{Hw99} and \citet{H01}.  Given the
higher energy resolution and bandpass of ASCA, it seems likely the
temperatures from this instrument are more reliable. Future
observations with Chandra and XMM-Newton should allow more accurate
measurements of the X-ray temperatures in hotter groups.

In Figure 6 we compare the metal abundances derived in our ROSAT survey
with abundances derived from ASCA observations. While there is considerable
scatter in the plot, both the ROSAT and ASCA data are consistent with 
sub-solar abundances in most cases. However, the reliability of metallicity 
measurements with these telescopes have been called into question
\citep{BB96,Buote99,Buote00}.  

The bolometric luminosity of each detected group was estimated from
the best-fit MEKAL model assuming the distance to the group given in
column 8 of Table 2.  To estimate the errors on the luminosities, we
varied the temperature and metallicity over their 90\% confidence
ranges.  For those groups without a metallicity measurement, we assume
a metallicity range from 0.0 to 1.0 solar.  The luminosities are
calculated out to the radius used in the spectral extractions. As this
radius is usually a fraction of the virial radius of each group, the
true bolometric luminosities for these groups could be substantially
higher \citep{HP00a,Mul00,H01}.  The bolometric luminosities for the
X-ray detected groups in our sample span four orders of magnitude
(from $\sim$ 10$^{40}$ to 10$^{44}$ h$_{\rm 100}$$^{-2}$ ergs
s$^{-1}$; see Figure 7).  A comparison of our estimated bolometric
luminosities with the common groups in \cite{HP00b} reveals good
agreement (see Figure 8).  We also compare our bolometric luminosities
with luminosities derived from ASCA observations (Figure 9). In
general, the ASCA luminosities are somewhat lower than the
ROSAT-derived luminosities. This result is not surprising given that
the ASCA luminosities are generally derived from smaller apertures.

Figure 10 plots the virial fraction versus the temperature of the
system for the X-ray detected groups in our sample (groups with poorly
constrained temperatures have been excluded). This plot indicates that
the majority of groups are detected to less than half of the virial
radius. There is also a trend for the hotter groups to be detected to
a larger fraction of the virial radius \citep{Mul00}. This trend is
important because it suggests that a larger fraction of the gas mass,
and thus, X-ray luminosity is detected in the hotter systems. This
trend continues to rich clusters of galaxies which are in general
detected out to approximately the virial radius \citep{Mul00}.

For the groups that were not detected, we estimate upper limits on the
X-ray luminosities using a method similar to that described in
\citet{M96a}. First, we measure the net counts (N) and rms error
($\sigma$) in a region 200 h$^{-1}$ kpc in radius centered on each
group (for a few nearby groups it was necessary to adopt a radius
smaller than 200 h$^{-1}$ kpc; see Table 2).  We then calculate the
flux a MEKAL plasma of temperature 1 keV and abundance 0.3 solar would
need to produce N+3$\sigma$ (if N$>$0) or 3$\sigma$ (if N$<$ 0)
counts. 

\section{Spatial Analysis}

For each group with a detected diffuse component we have fit the
two-dimensional surface brightness profile using a modified King
function (the so-called `$\beta$-model'):

\centerline{S(R)=S$_{\rm o}$
(1.0 + (R/R$_{\rm core}$)$^2$)$^{-3\beta + 0.5}$.}

\noindent{The model is
first convolved with the 1 keV ROSAT PSPC point spread function, and
then fit to the data with S$_{\rm o}$, R$_{\rm core}$, $\beta$ and 
the position of the center of the emission
as free parameters. Following \cite{HP00a}, we also fit both circular
and elliptical fits. In the latter fits, the ellipticity and position angle
are additional free parameters. The results of the elliptical fits are
given in Table 3.
During the fits, each free parameter was allowed to vary between a 
reasonable lower and upper bound. For a large number of groups, the 
best-fit core radius was at the lower bound value (0.1$'$). This value
is smaller than the resolution of the ROSAT PSPC. In these cases, 
the core radius is listed as `$<$0.1' in Table 3. To determine what
effect this lower bound might have on the other derived parameters, we
fit a number of groups again without the lower bound. These fits suggest
that removing the lower bound on R$_{\rm core}$ has a minimal impact on
the derived values of $\beta$ and ellipticity (see also the similar finding
of \cite{HP00a}). 

The goodness of each fit was estimated using $\chi$-squared statistics.
Because the number of counts in each bin is small, we use the 
prescription of \cite{G86} to calculate the standard deviation.
While the single-$\beta$ models fits provide an adequate description
of the images in some cases, in general the fits are poor. 
Previous studies of X-ray luminous groups indicate that the surface
brightness profiles of groups are complicated, often requiring
multiple components to adequately describe the data
\citep{MZ98,HP00a}. In particular, single King models tend to
underestimate the flux at the very centers of luminous groups.
\cite{MZ98} suggested that groups contain two components, one
associated with the central galaxy (the `central' component) 
and a second component associated
with the group as a whole (the `extended' component).  
For those groups in our sample with a sufficient 
number of counts ($\sim$ 1000), we fit the two-dimensional surface
brightness profiles with a 
two-component $\beta$ model. Initially, we experimented with allowing
both components to be elliptical in shape (i.e. let the ellipticity and
position angle be free parameters). We found, however, that it was 
not possible to constrain these fits in most cases. Next, we adapted
the proceedure used by \cite{HP00a} and allowed one of the components
to be elliptical and required the other to be circular. While these
models provided adequate fits to some groups, it was often necessary to
restrict the number of free parameters even further in some cases.
In particular, the $\beta$ values for the circular models often reached
unphysically high values. When this occurred, we fit the surface
brightness profiles again with the $\beta$ value of the core component 
fixed at a value of 1.0 (\cite{HP00a} encountered a similar problem
in their fits; Stephen Helsdon, private communication). 
In those cases where a double $\beta$-model fit
was performed, the results are given in Table 3. 
Note that for the double $\beta$-model fits, the elliptical component
is always listed first (i.e. the elliptical component is 
identified as the `extended component'). 
In a few cases, this results in the 
core radius of the `extended component' being smaller than the 
core radius of the `central component'. 

Figure 11 shows the distribution of extended component $\beta$ values 
for our group sample (we adopt the two component fits where 
available). The mean value of $\beta$ for the sample is 0.47$\pm{0.16}$.
This value is in excellent agreement with the weighted mean that 
\cite{HP00a} find for their sample (0.46$\pm{0.06}$) and is somewhat lower
than the typical value found for rich clusters \citep{AE99,MME99}.

\section{Detection Statistics}

The existence of a hot intragroup medium in poor groups of galaxies
has been firmly established over the past decade. However, the
frequency of diffuse X-ray emission in groups is still rather
uncertain. Most previous studies using ROSAT `pointed mode' data have
been restricted to rather small samples selected in a potentially
biased way.  For example, many of the targets in these surveys were a
priori known to be bright X-ray sources based on previous detection
(i.e. in the ROSAT All-Sky Survey). Furthermore, a significant
fraction of the groups studied to date come from the Hickson Compact
Group catalog, which may not be representative of groups in general
(see, however \citet{HP00b}). \citet{Ma00} attempted to overcome these
problems by looking for X-ray emission in the ROSAT All-Sky Survey
data for a large sample of groups selected from the CfA redshift
survey. Correcting for selection effects, \citet{Ma00} estimate that
approximately 40\% of groups are extended X-ray sources.

Our present survey represents the largest sample of groups studied to
date with ROSAT PSPC `pointed mode' data. Diffuse emission is detected
in at least 61 of the 109 groups (56\%) in our sample (not counting
the 6 questionable detections) down to the detection limits of our
survey. Unfortunately, our sample is plagued
by the same biasing issues that effect the earlier pointed mode
studies. 
A deep X-ray survey of a large, optically-selected group sample should
be a priority for the current generation of X-ray telescopes.

Early studies of groups with ROSAT suggested that X-ray emission is
largely restricted to systems with low spiral fractions
\citep{E94,PBE95,M96a}.  However, \citet{P96} found that some
spiral-rich groups do indeed contain X-ray emission. Figure 12 shows
the distribution of spiral-fractions for our entire sample and for the
X-ray detected groups. This figure is consistent with \citet{P96}'s
suggestion that some spiral-rich groups are X-ray
detected. Interestingly, however, all of the groups with a detectable
intragroup medium contain at least one early-type galaxy.  Therefore,
the ROSAT data are consistent with no spiral-only groups containing a
detectable hot intragroup medium.

Several explanations for the failure to detect spiral groups in X-rays
have been proposed.  One possibility is that spiral-only groups are
not real, physical systems, but rather chance superpositions of
galaxies along the sight.  However, our own Local Group would be
defined as \lq\lq spiral-only\rq\rq\ if it was placed at the distance
typical of the groups in our survey \citep{ZM98}. Therefore, this
possibility seems unlikely.  Another possibility is that the gas
density is significantly lower in spiral-only groups. Lower gas
density in spiral groups may in fact be consistent with expectations
from recent preheating models \citep{P99}.  
The temperature of the gas in spiral-only groups may also
be too cool to produce detectable X-ray emission \citep{M96b}.
 Based on their
velocity dispersions, the virial temperatures of spiral-only groups do
tend to be lower than those of their early-type dominated
counterparts. A \lq\lq warm\rq\rq \ intragroup medium might produce
detectable features in the far-ultraviolet or X-ray spectra of
background quasars \citep{M96b,PL98,H98}.  High-ionization absorption
line features consistent with cooler groups have recently been
reported. For example, \citet{F01} have reported the detection
of an O VIII Ly$\alpha$
absorption line associated with a spiral group along the sightline
towards PKS 2155-304. Similarly, some of the O VI systems found with
HST and FUSE may be consistent with an origin in small galaxy groups
\citep{S02}. Further absorption studies should provide insight into
the presence of warm intragroup gas in spiral-only groups.

\section{Summary}

We have searched for diffuse X-ray emission in 109 nearby galaxy
groups using data taken with the ROSAT PSPC during its `pointed mode'
phase. We find evidence for a hot intragroup medium in approximately
half of the groups in our sample. Although we detect many spiral-rich
groups, none of the spiral-only groups in our survey show clear
evidence for the presence of diffuse X-ray emission. Spiral-only
groups may contain intragroup gas, but the temperature or gas density
may be too low to produce appreciable X-ray emission.

In general, nearby groups are detected to only a fraction of the
virial radius with the ROSAT PSPC. Therefore, a significant amount of
the gas mass likely occurs beyond the current X-ray detection
radius. Future observations with Chandra and XMM-Newton should provide
further insight into the distribution of hot gas in these 
cosmologically-important systems.

\acknowledgments

The authors acknowledge useful conversations with
Alex Athey, Steven Helsdon,
Donald Horner, Trevor Ponman, Ann Zabludoff and Marc Zimer.  We also
acknowledge Mark Donikian for help with the figures.  This project
made extensive use of NASA's HEASARC, NED and Skyview databases.
Partial support for this project was provided by NASA grant NAG 5-3529.

\begin{figure}
\caption{Contour maps of diffuse X-ray emission overlayed on the STScI
digitized sky survey for each group detected in our survey.  The
coordinates are in epoch J2000.  Emission from point sources in the
field has been removed except for emission associated with the central
galaxy.  The contours correspond to 3$\sigma$, 5$\sigma$, 10$\sigma$,
20$\sigma$ and 40$\sigma$ above the background. The data have been
smoothed with a Gaussian of width 30$''$. 
The complete Figure 1 is available at: http://www.ociw.edu/$\backsim$mulchaey/Atlas/atlas.html}
\clearpage
\plotone{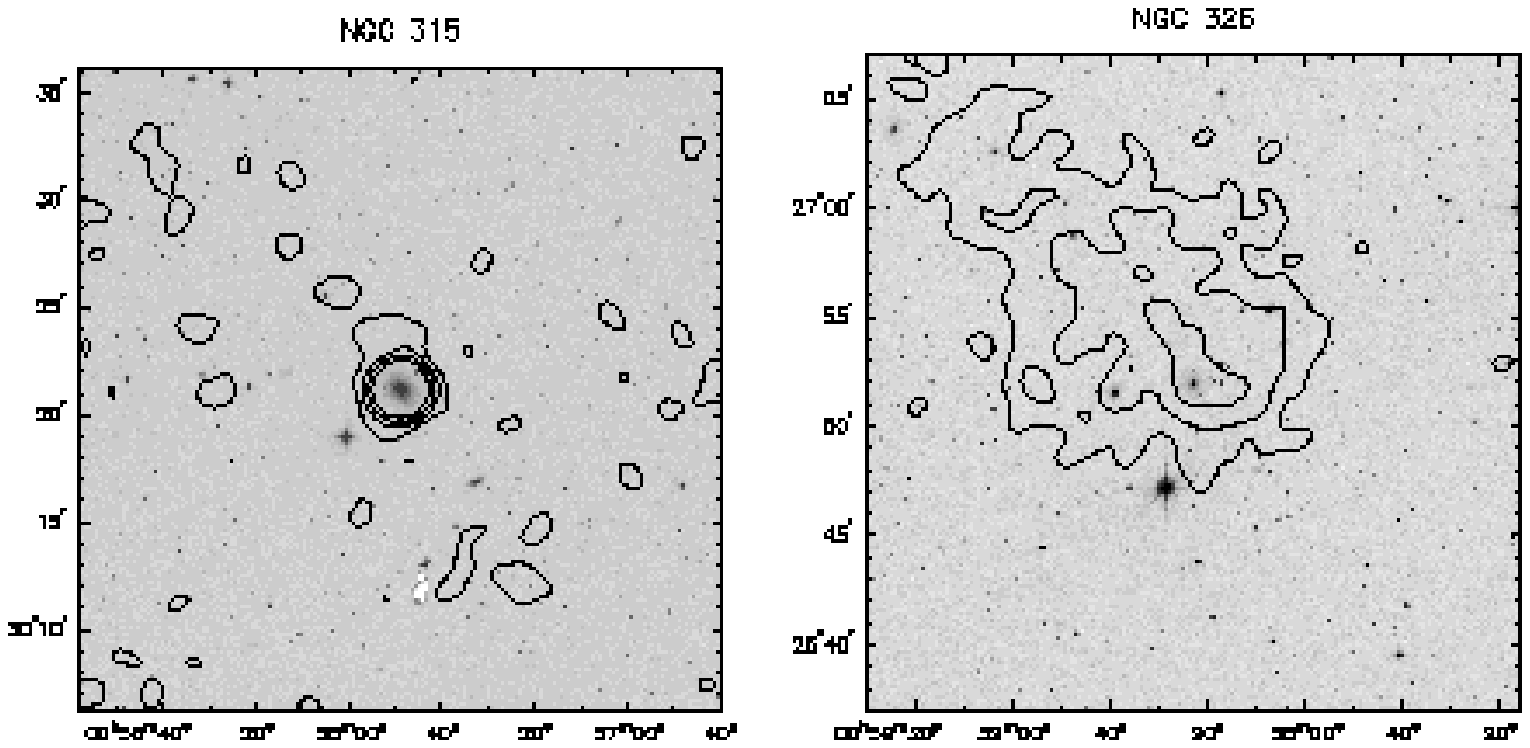}
\end{figure}

\begin{figure}
\plotone{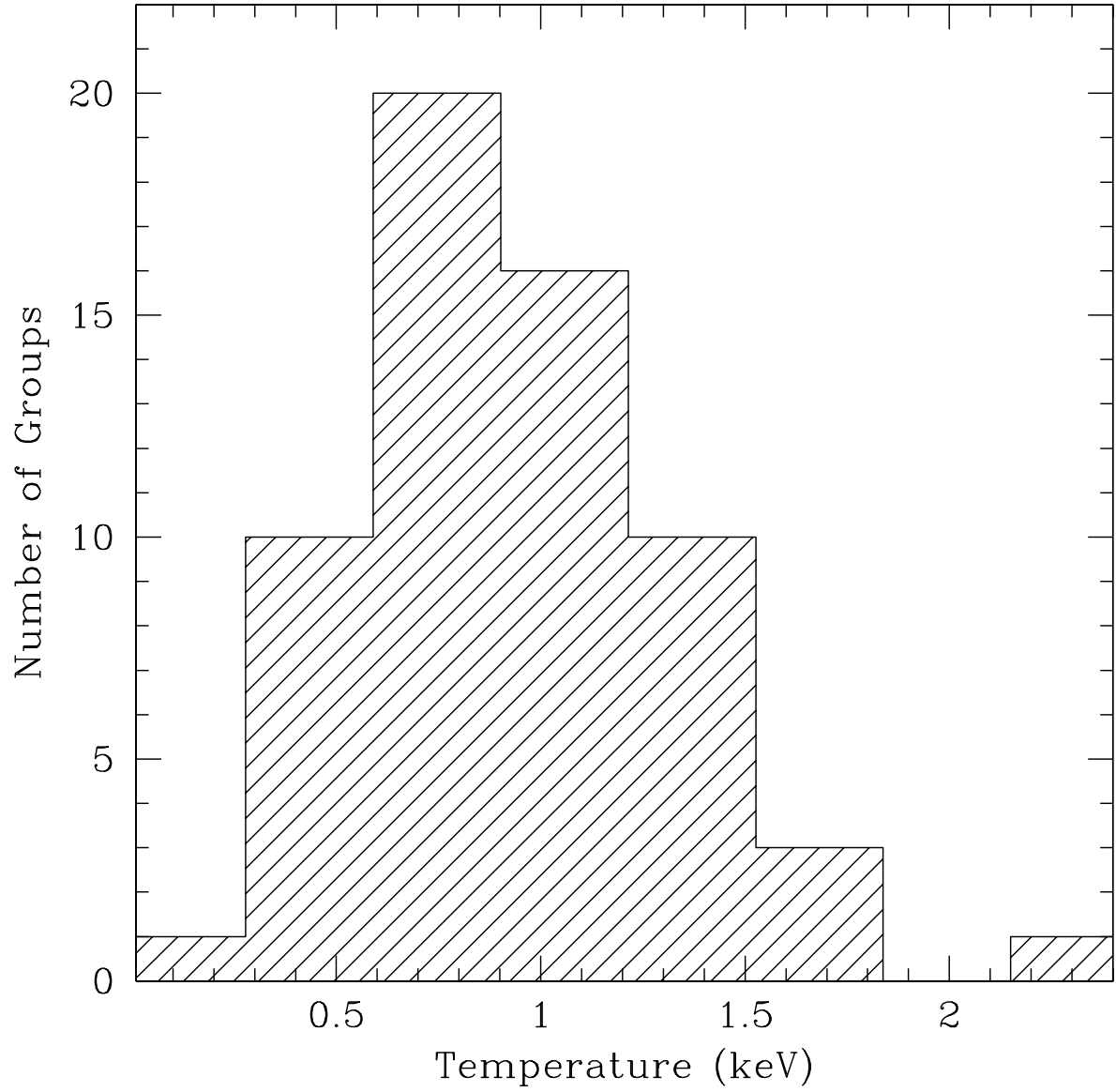}
\caption{Distribution of derived gas temperatures for the group sample.} 
\end{figure}

\clearpage

\begin{figure}
\plotone{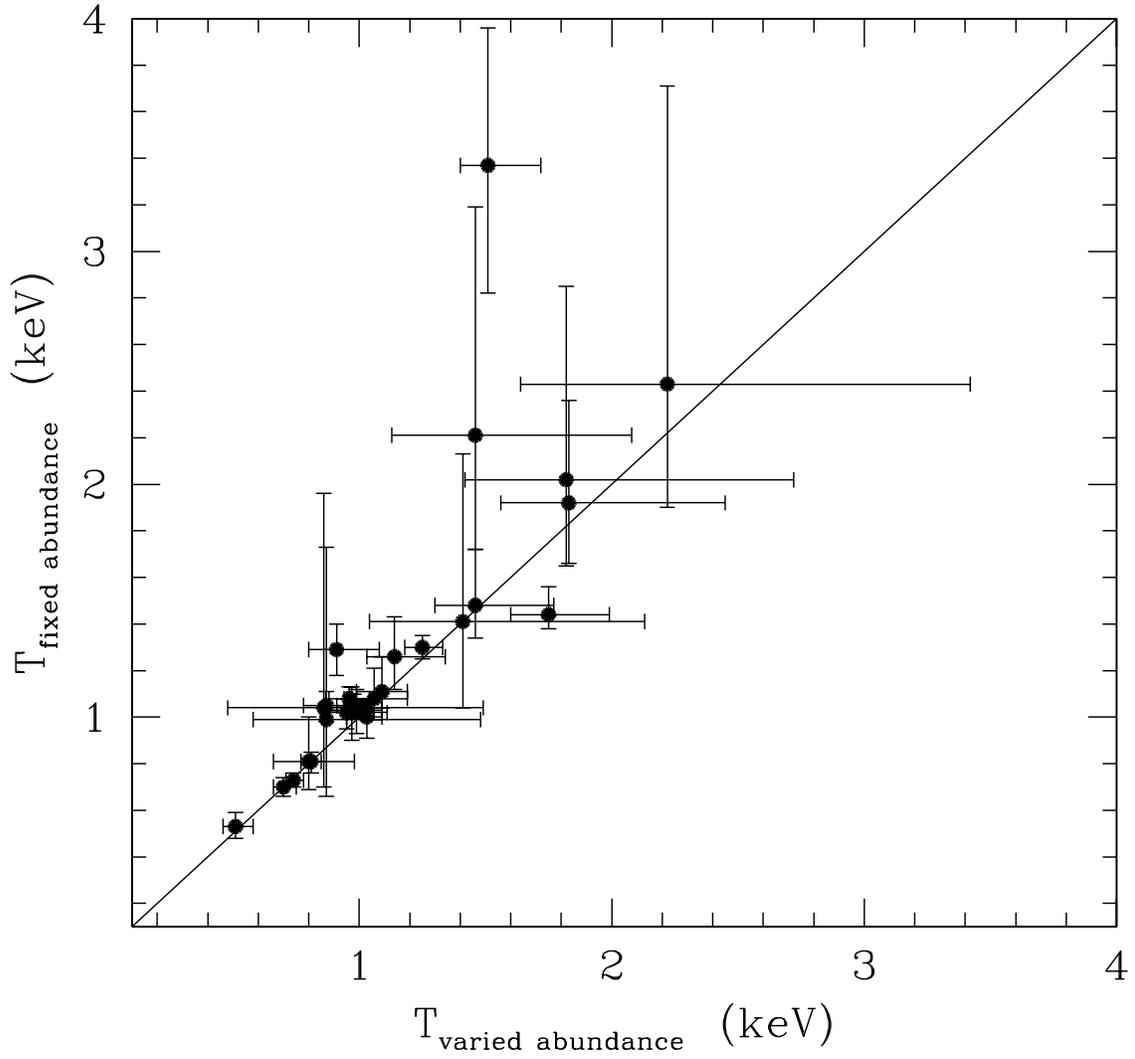}
\caption{A comparison of the ROSAT derived temperatures for a MekaL
model with the metallicity allowed to vary versus a MekaL model with
the metallicity fixed to 0.3 solar. }
\end{figure}

\begin{figure}
\plotone{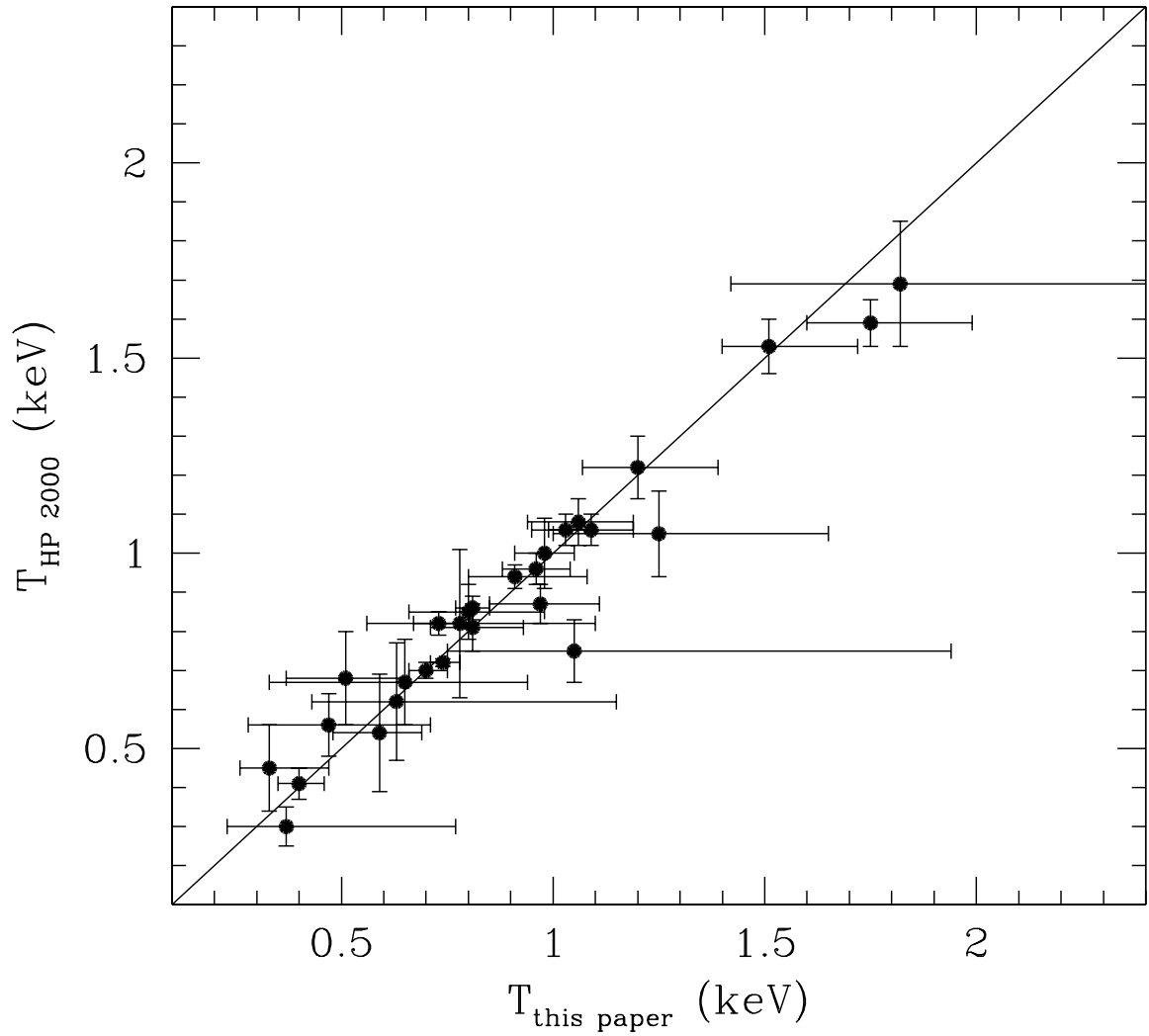}
\caption{A comparison of derived gas temperatures from the present
survey with those in Helsdon \& Ponman (2000b). Overall, the agreement
is very good.}
\end{figure}

\begin{figure}
\plotone{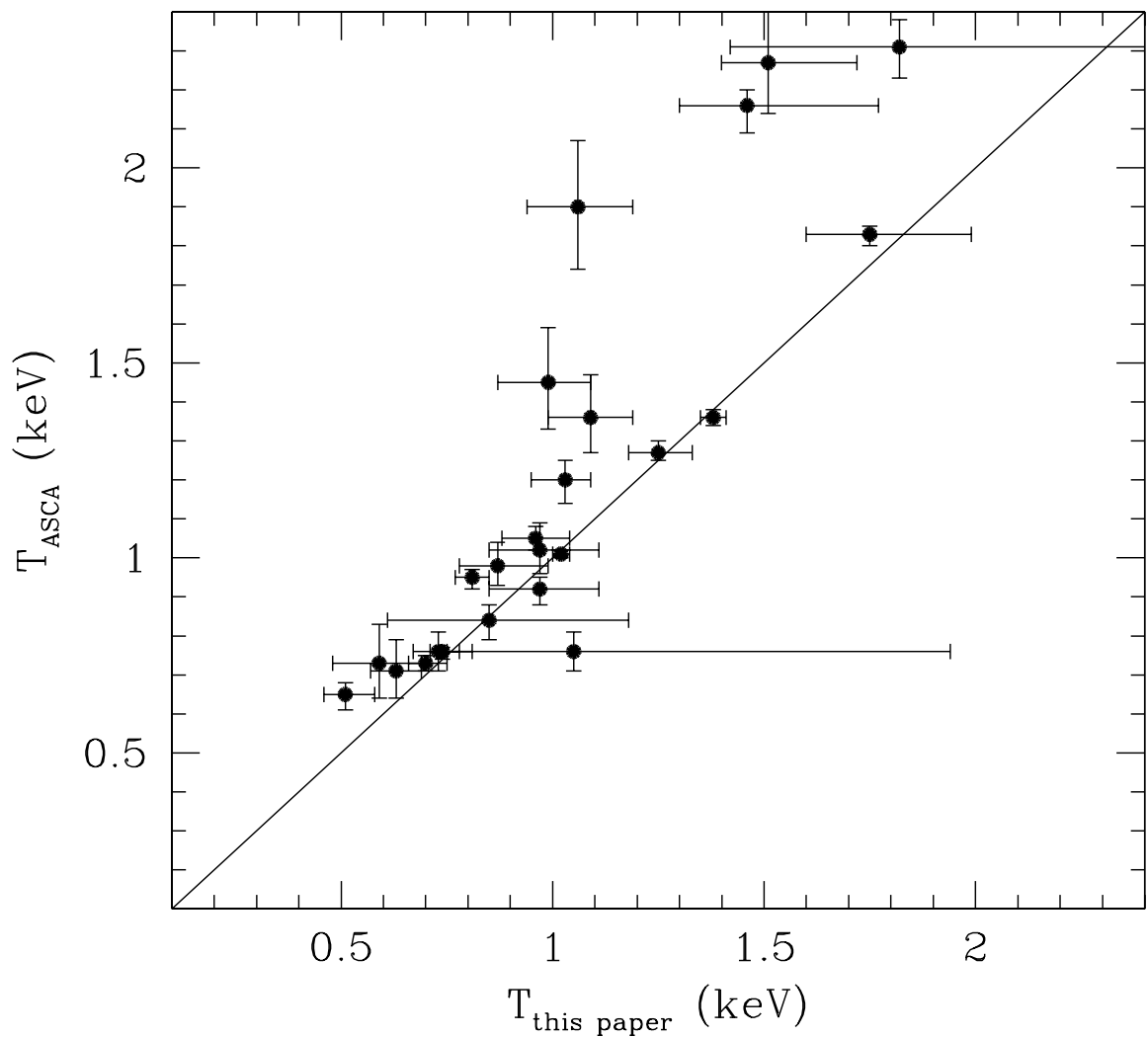}
\caption{A comparison of the best-fit ROSAT derived temperatures with
temperature measurements made with the ASCA X-ray telescope. The ASCA
measurements are taken from the literature and assume an isothermal
MekaL model. There is a trend for the
ROSAT data to underestimate the gas temperature especially for
temperatures greater than 1 keV.}
\end{figure}

\begin{figure}
\plotone{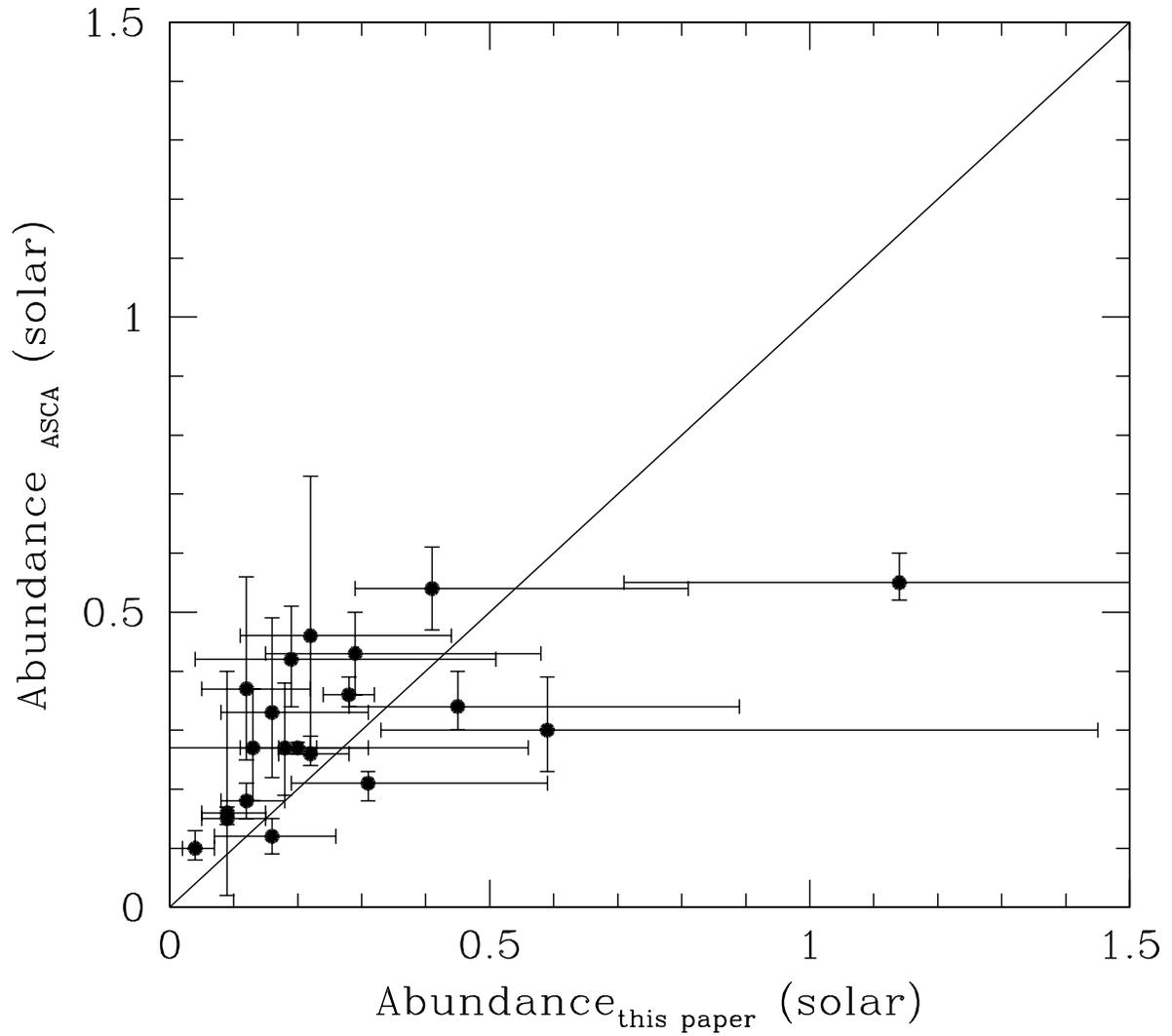}
\caption{A comparison of the best-fit ROSAT derived abundances with
abundance measurements made with the ASCA X-ray telescope. The ASCA
measurements are taken from the literature and assume an isothermal
MekaL model.} 
\end{figure}

\begin{figure}
\plotone{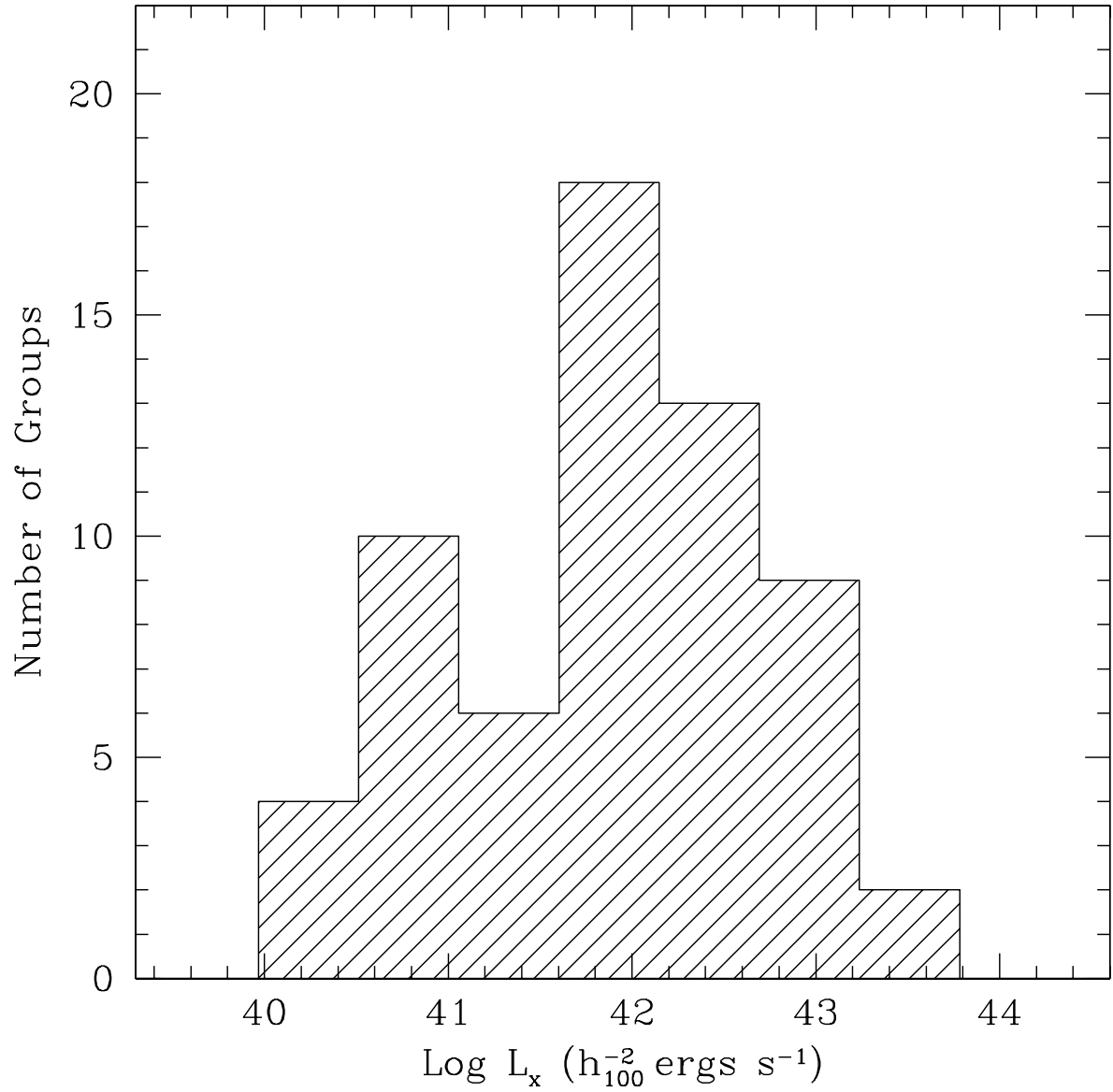}
\caption{Distribution of bolometric luminosities for the groups in the
sample with detected extended X-ray emission.}
\end{figure}

\begin{figure}
\plotone{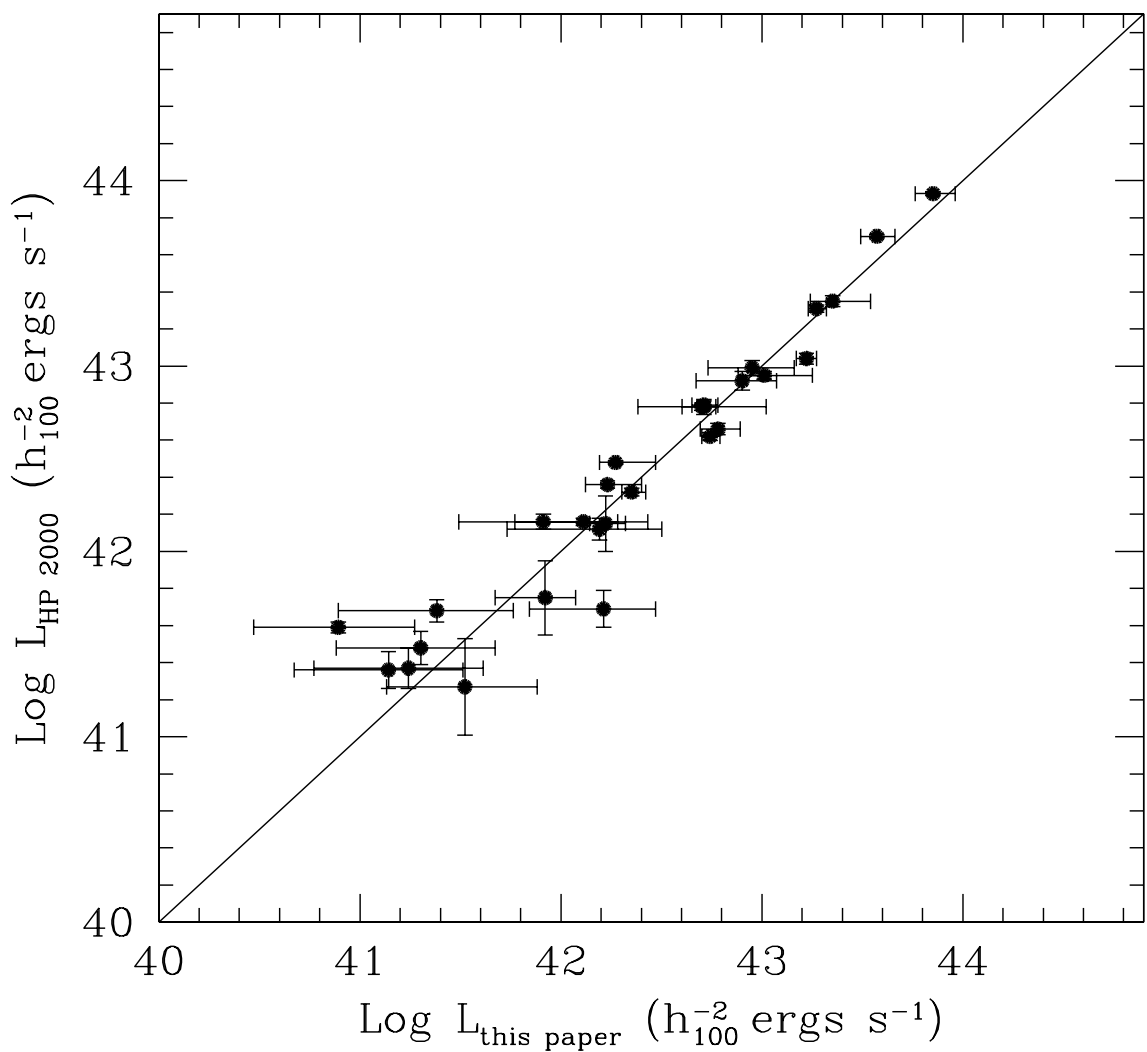}
\caption{A comparison of bolometric luminosities from the present
survey with those in Helsdon \& Ponman (2000b). Overall, the agreement
is very good.}
\end{figure}

\begin{figure}
\plotone{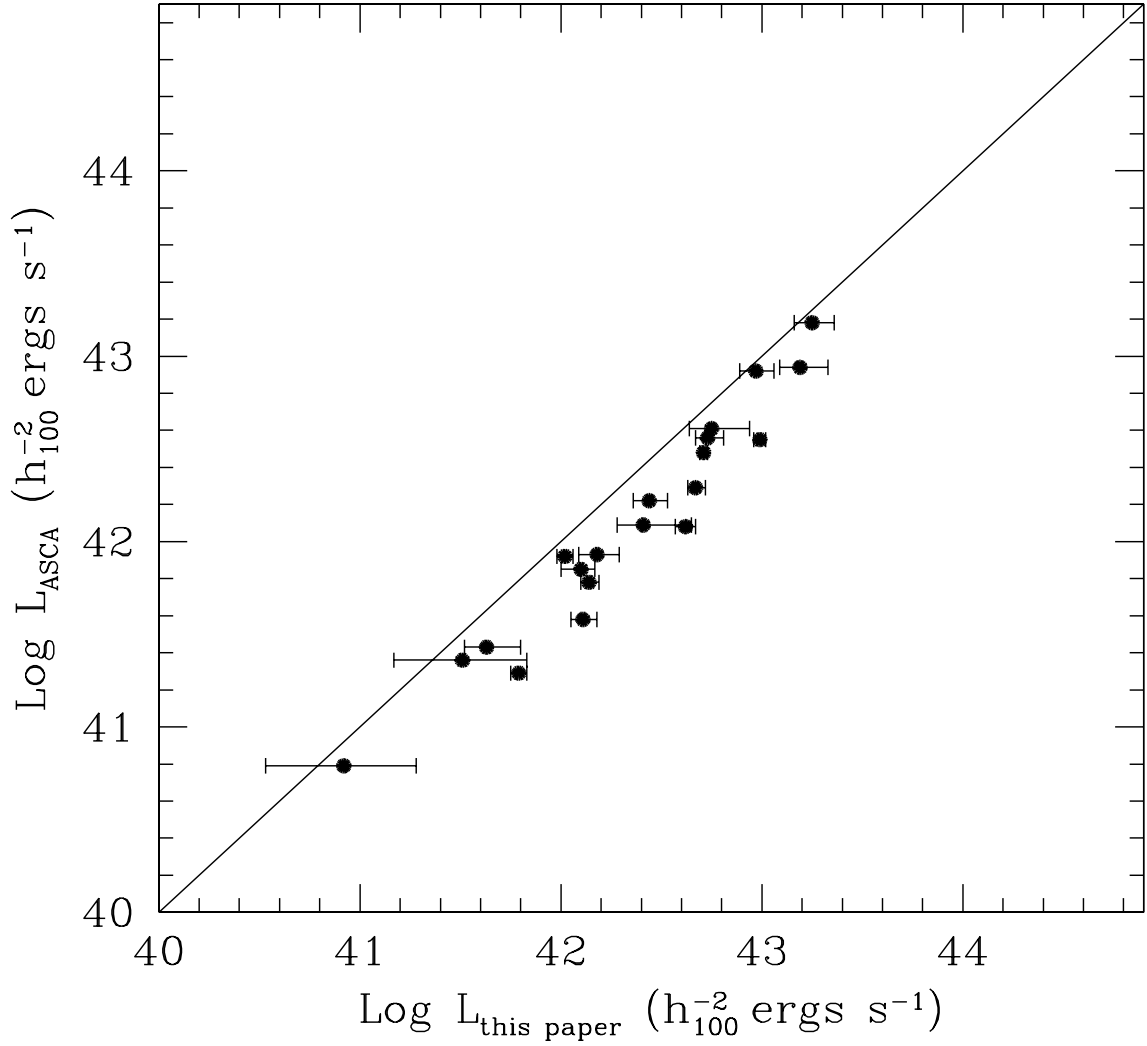}
\caption{A comparison of bolometric luminosities derived from the present
survey with bolometric luminosities derived from 
ASCA observations. The ASCA luminosities tend to be smaller than those derived
from the ROSAT data. This is most likely because smaller extraction radii 
were used in the ASCA analysis.}
\end{figure}

\begin{figure}
\plotone{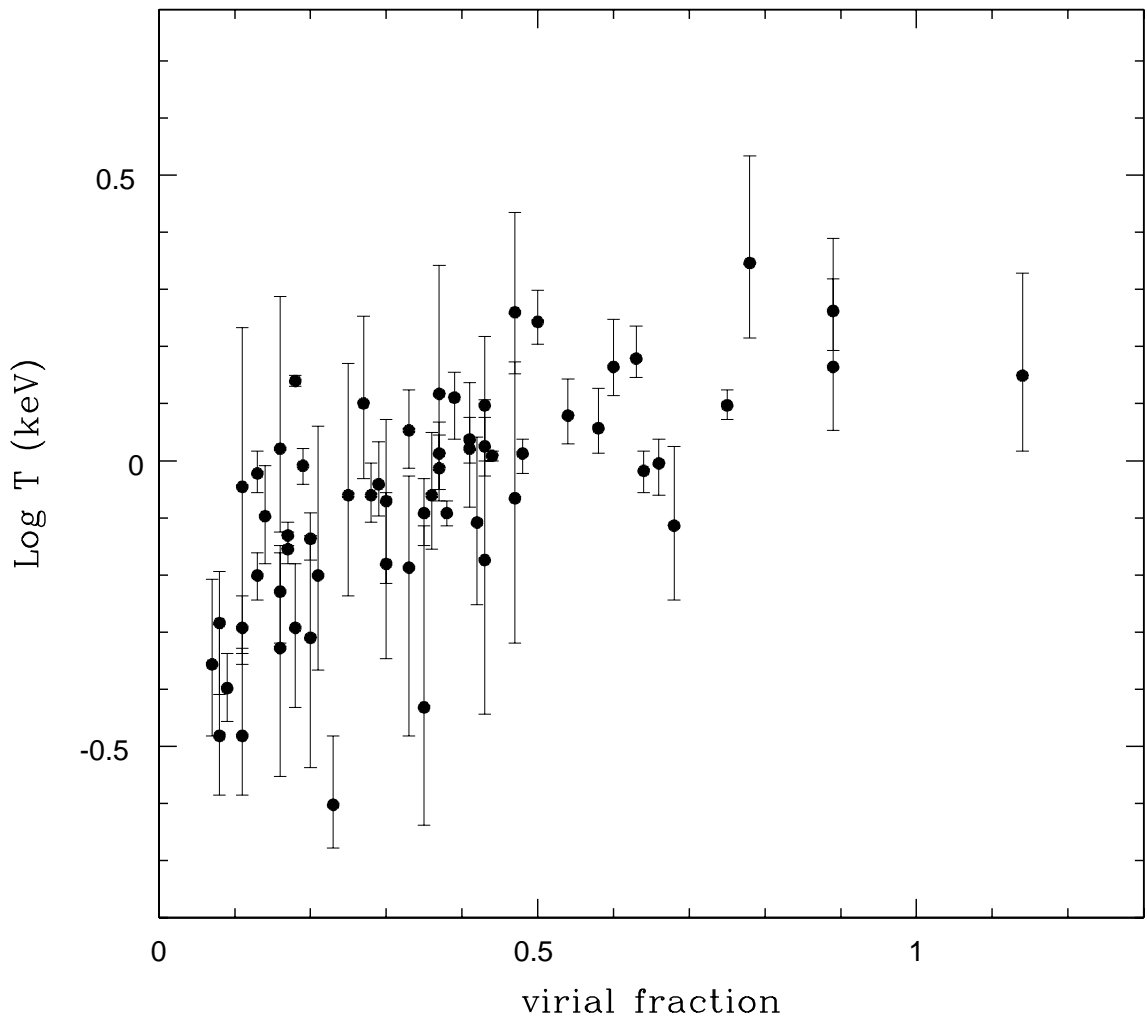}
\caption{Total radius of X-ray extent plotted as a fraction of the
virial radius of each system versus the logarithm of the temperature
for the group sample.  The virial radius for each system was
calculated assuming r$_{\rm virial}(T)$ = 1.85 (T/10keV)$^{0.5}$
(1+z)$^{-1.5}$ h$_{\rm 100}^{-1}$ Mpc (Evrard et al 1996).}
\end{figure}

\begin{figure}
\plotone{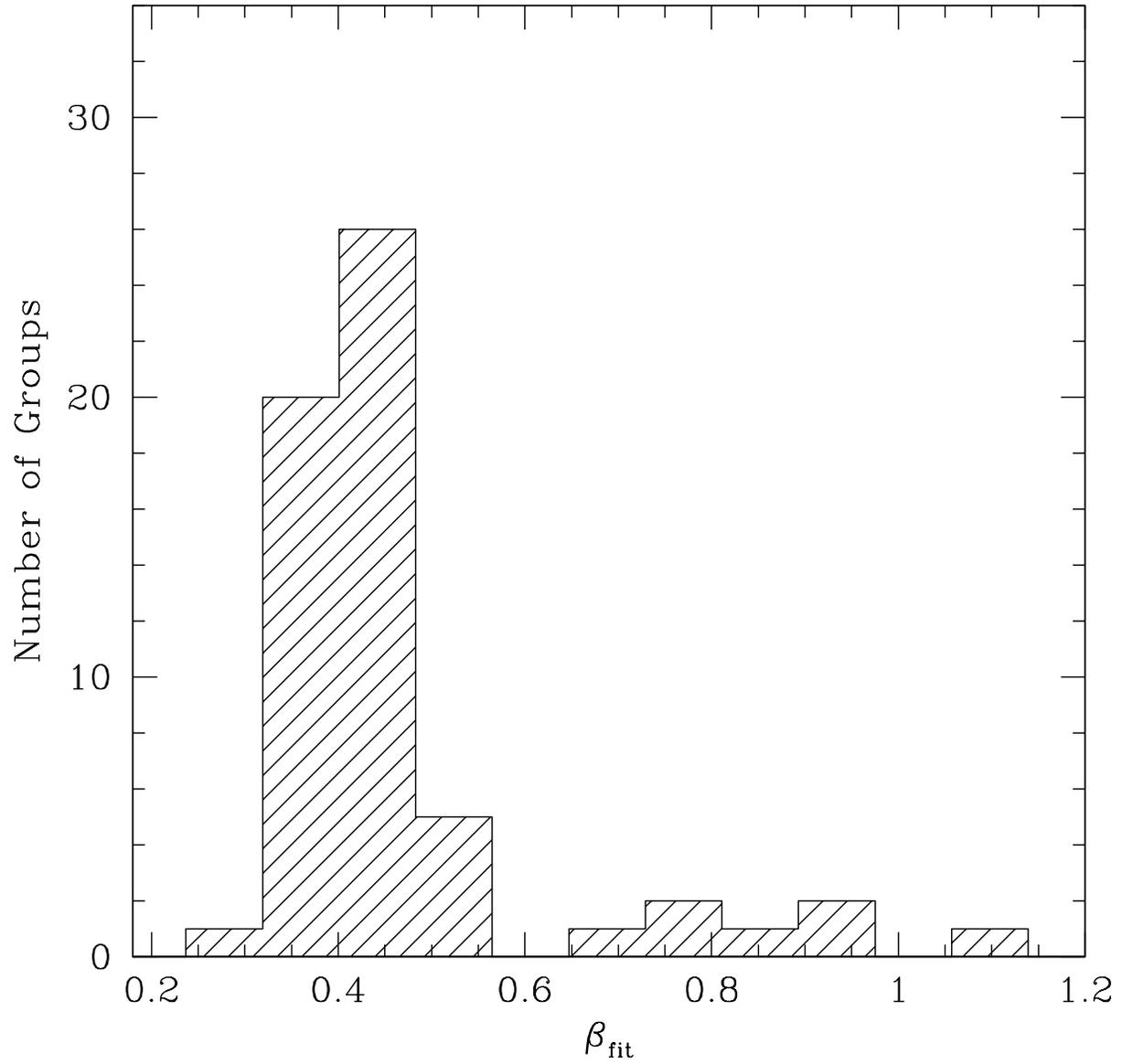}
\caption{Distribution of $\beta$ values for the extended component derived
from $\beta$-model fits. The $\beta$ values are taken from the two component
fits where available, otherwise the single component value is used.}
\end{figure}

\begin{figure}
\plotone{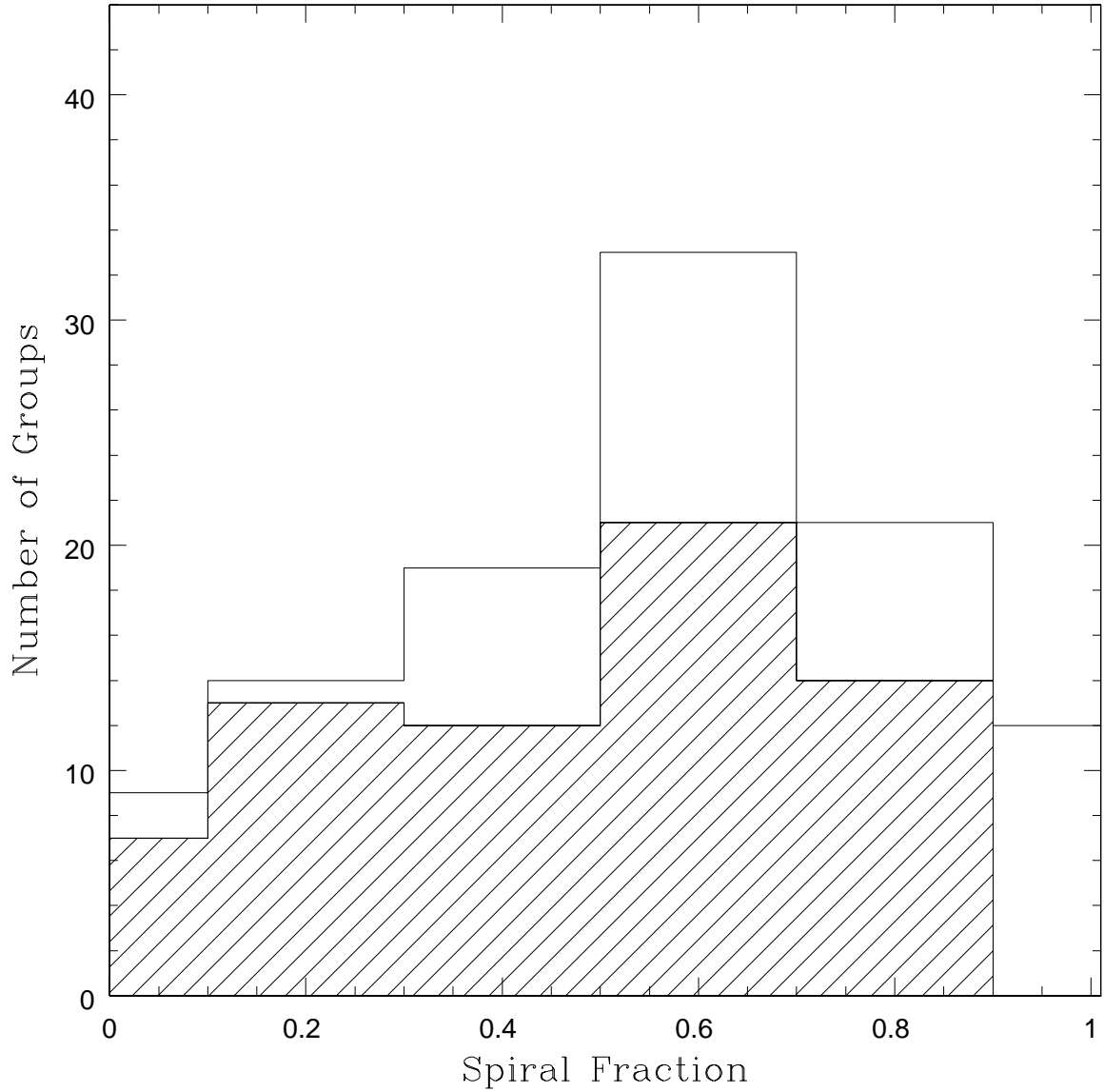}
\caption{Histogram of spiral fraction for the entire group sample
(open) and for the X-ray detected groups (filled). The 6 marginally
detected groups are included in the X-ray detected group
histogram. Note that no spiral-only group is detected in our survey.}
\end{figure} 

\input{table1.tex}

\clearpage

\input{table2.tex}

\clearpage

\input{table3.tex}




\end{document}

%% file: table1.tex

	\begin{deluxetable}{lcccccccl}
	\tablecolumns{9}
	\tablewidth{0pc}
	\tablecaption{Group Sample}
	\tablehead{
	\colhead{Group} & \colhead{RA} & \colhead{Dec} & 
	\colhead{N$_{\rm gal}$} & \colhead{z} & 
	\colhead{$\sigma$} &
	 \colhead{f$_{\rm spiral}$} & \colhead{ref} & \colhead{Catalog} \\
	 &\colhead{(J2000)}& \colhead{(J2000)}&  & \colhead{} & \colhead{km/s} &
	 & & \colhead{Name$^1$}}  \startdata
	NGC 7805& 00 02 28.0 & +31 28 42.1& 3 & 0.0164& 71 & 0.66 & 1 & G1/GH176/N173 \\ 
	NGC 43  & 00 13 05.8 & +30 58 40.8& 3 & 0.0160& 63 & 0.33 & 1 & G1/N1 \\
	HCG 2    & 00 31 30.0& +08 25 52.5& 3 & 0.0144& 55 & 1.00 & 2,3 & H2 \\
	HCG 4    & 00 34 16.0 & -21 26 48.0& 5 & 0.0245& 775& 0.66 & 4,5 & H4 \\
	HCG 3    & 00 34 27.5 & -07 35 34.2& 3 & 0.0255& 251& 0.33 & 2,3 & H3\\      
	IC 1559 & 00 36 38.3& +23 58 30.2& 3 & 0.0158& 470& 0.33 & 6 & GH5 \\
NGC 315 & 00 57 48.9& +30 21 08.8 & 4 & 0.0164& 122 & 0.25 & 1 & G14/GH8/N6 \\	
NGC 326  & 00 58 22.3 & +26 52 03.7& 9 & 0.0477&674 &0.00 & 5,7&-  \\
	NGC 383 & 01 07 25.0& +32 24 45.2& 40 &  0.0173& 567 & 0.50 & 8,6& G18/GH9  \\
	NGC 491 & 01 20 42.3& -34 01 17.3& 6 & 0.0126& 86& 1.00 & 9& MDL58  \\
NGC 507  & 01 23 40.0 & +33 15 20.0& 21 & 0.0170& 179 &0.82 & 10 & G26\\	
NGC 524 & 01 24 47.8& +09 32 19.0& 8 & 0.0083& 205& 0.50 & 1 &G23/GH13/N11\\
	NGC 533  & 01 25 31.4& +01 45 32.8& 36 & 0.0181&464 &0.39 & 9& GH14\\  
	HCG 10   & 01 26 07.4 & +34 41 27.4& 4 & 0.0161& 209& 0.75 & 2,3 & G26/H10 \\
	HCG 12   & 01 27 33.2& -04 40 40.3& 5 & 0.0485&240& 0.20 & 2,3 &H12 \\
	NGC 584 & 01 33 21.5& -07 01 16.5& 4 & 0.0066& 107& 0.63& 10 &G27/HG45  \\
	NGC 664  & 01 44 02.7 & +04 19 02& 6 & 0.0180& 130&  1.00& 9 & - \\
	NGC 720 & 01 53 00.5& -13 44 18.4 & 4 & 0.0059& 161&  0.75& 10 & G38  \\
	NGC 741  & 01 56 21.0& +05 37 44.2& 41& 0.0185& 432  & 0.44  & 9& N20\\
	HCG 15   & 02 07 37.6& +02 10 40.8& 6 & 0.0228& 427& 0.50 & 2,3 & H15 \\
	HCG 16   & 02 09 27.1& -10 07 37.2& 9 & 0.0131& 85& 0.86 & 8,4,5 & G49/H16 \\ 
	UGC 1651 & 02 09 38.2& +35 47 49.8& 3 & 0.0363& 111 & 0.00 & 5 & - \\
        HCG 18   & 02 39 06.8& +18 22 59.1& 3 & 0.0137&  31 & 1.00 & 5 & H18 \\
	NGC 1044 & 02 41 06.3& +08 44 18.2& 13 & 0.0205& 325& 0.66& 5 & - \\
        IC 1860  & 02 49 33.4& -31 11 27.1& 28 & 0.0231 & 452& 0.32 & 5 & MDL67 \\	
HCG 22   & 03 03 31.3& -15 40 32.5& 8 & 0.0090& 183& 0.75 & 4,5,8 & G81/H22 \\
        IC 1880  & 03 06 28.4&-09 43 42.5& 7 & 0.0338& 208 & 0.50& 5 & - \\
	HCG 23   & 03 07 06.5& -09 35 07.7& 14 & 0.0152& 339& 0.40 & 8,4,5 & H23 \\
	HCG 26   & 03 21 54.2 & -13 38 45.1& 7 & 0.0316& 200& 0.57 & 2,3 & H26 \\
	NGC 1332 & 03 26 17.1& -21 20 05.1& 6 & 0.0052& 347 & 0.50 & 5 & G97/HG32 \\
	UGC 2755 & 03 29 24.0& +39 47 35.2& 5 & 0.0230& 406 & 0.66 & 5& - \\
        NGC 1399 & 03 38 29.3& -35 27 02.3& 55& 0.0036& 312& 0.55 &11 & G96/HG17/MDL52 \\
        NGC 1407& 03 40 16.4& -18 35 41.7& 8 & 0.0057& 145&  0.25& 10 & G100/HG32 \\
	NGC 1587& 04 30 40.6& +00 40 00.6& 4 & 0.0122& 106& 0.50 & 1 &G117/N33 \\
	HCG 31   & 05 01 40.3& -01 15 24.3& 11 & 0.0131& 141 & 1.00 &8,2,3 & G123/H31 \\
	HCG 33   & 05 10 47.9& +18 02 04.7& 4 & 0.0260& 155& 0.25 & 2,3 & H33\\
	NGC 1961& 05 43 51.8& +69 18 09.4& 4 & 0.0139& 139& 1.00 &  10 & G132  \\ 
	NGC 2300 & 07 32 14.2& +85 42 33.7& 13 & 0.0069&  228& 0.66 & 5,8 &G145/HG92 \\
	NGC 2484 & 07 58 28.2 & +37 47 12.9&  2 & 0.0415& - & 0.00 & 5 & - \\
	NGC 2563& 08 20 35.7& +21 04 03.9& 44 &  0.0159& 419& 0.72 & 12 & G158 \\
	HCG 35   & 08 45 19.0& +44 30 54.7& 6 & 0.0542&  316& 0.20 & 2,3 & H35 \\
	NGC 2769& 09 10 22.8& +50 23 45.3& 3 & 0.0166& 125& 1.00 & 1 & G168/N35 \\
	HCG 37& 09 13 47.3& +30 00 37.1& 5 & 0.0223& 398& 0.40 &2,3& GH42/H37 \\
	NGC 2805& 09 23 12.8& +63 47 09.0& 4 & 0.0055& 69 & 0.50 & 1 &G186/GH46/N36  \\
	HCG 38   & 09 27 38.9& +12 16 50.9& 3 & 0.0292& 13 & 1.00 & 2,3 & H38  \\
	HCG 40   & 09 38 54.5 & -04 51 05.1& 7 & 0.0221& 280 & 0.50 & 4,5 & H40 \\
	HCG 42   & 10 00 14.3& -19 38 13.3& 35 & 0.0128& 266& 0.57 & 12 & G186/HG37/H42 \\
	HCG 44   & 10 18 00.5& +21 48 44.1 & 4 & 0.0046&135 & 0.50 & 2,3 & G194/H44/N43 \\
	MKW 2& 10 30 10.4& -03 09 55.0& 33 & 0.0365& 603 & 0.20 & 5 & - \\
	HCG 48   & 10 37 50.0 & -27 07 17.5& 3 & 0.0094& 302 & 0.33 & 2,3 & G210/H48 \\
	CGCG154-041& 10 43 18.7& +31 31 07.0& 4 & 0.0350& 312 & - & 5 & - \\
	NGC 3396& 10 50 41.1 & +33 13 17.3& 6 & 0.0053&73&1.00&10 &G218/GH67/HG65/N 55 \\
	NGC 3557& 11 09 57.4& -37 32 17.1& 22 &0.0095& 282 &  0.63 &5,12& G229 \\
NGC 3607 & 11 16 54.1& +18 03 18.9& 7 & 0.0037& 92 & 0.86 & 1 & N66 \\
	HCG 50   & 11 17 06.1 & +54 55 06.7& 5 & 0.1391& 468 & 0.00 & 2,3 & H50 \\
	NGC 3647 & 11 21 45.8& +02 51 40.0& 6 & 0.0495& 383 & 0.00 & 5 & - \\ 
	NGC 3665& 11 24 43.5& +38 45 43.4& 4 & 0.0069& 29 & 0.25 & 1 & G236/GH79/N68  \\
	HCG 57   & 11 37 52.9 & +21 58 17.8& 7 & 0.0304& 269& 0.43 & 2,3 & H57 \\
	HCG 58  & 11 42 10.8& +10 18 24.0& 5 & 0.0206& 215 & 0.60 & 1 & H58/N82 \\
	NGC 3923 & 11 51 01.6& -28 48 19.9& 5 & 0.0046& 98 & 0.40 & 10 & G255 \\
NGC 4065& 12 04 09.1 & +20 17 06.0& 7 & 0.0232& 482 & 0.43 & 1 & N91 \\
	NGC 4073 & 12 04 27.4 & +01 53 42.4& 19 & 0.0201& 698 & 0.22 & 5& -  \\ 
	NGC 4104 & 12 06 45.4 & +28 10 11.9& 8 & 0.0286& 423 & 0.14 & 5& -  \\
	NGC 4125& 12 08 06.6 & +65 10 28.9 & 3 & 0.0050& 58  & 0.66& 10 & G274/GH94 \\
	NGC 4168& 12 13 38.9& +13 01 19.3 & 4 & 0.0077& 152& 0.50 & 1 & N98 \\
NGC 4261& 12 19 23.5& +05 49 36.4 &8&0.0068&108 &0.50&10&G278/GH106/HG41/N99
\\	
SHK 202  & 12 19 44.2& +28 26 35.2& 5 & 0.0267& 600 & 0.00 & 5& - \\
	NGC 4278 & 12 20 06.8& +29 16 50.7& 17 &0.0032&191&0.71 &10& G279/GH94/HG60/N100 \\
	NGC 4357& 12 20 51.1& +49 18 21.7& 3 & 0.0142& 273& 0.66 &  6 &GH103  \\
	NGC 4325 & 12 23 06.6& +10 37 13.2& 26 & 0.0254&  328& 0.66 & 12& -  \\
	NGC 4291& 12 24 41.2& +75 19 56.4& 11&0.0068& 132&0.73& 10 &G284/GH107/HG88 \\
	NGC 4615& 12 41 16.0& +26 13 33.1& 3 & 0.0158& 47 & 0.33 & 1 & N108  \\
	NGC 4636 & 12 42 49.8& +02 41 14.3& 12 & 0.0044& 463 & 0.83 & 1 & N104 \\
	HCG 62   & 12 53 06.1& -09 12 16.3& 63 & 0.0145& 390 & 0.47 & 12& H62 \\
	NGC 5044& 13 15 26.1& -16 23 01.7& 9 & 0.0082& 119 & 0.56 & 10 & G338   \\
	NGC 5101& 13 18 57.0& -26 53 39.5& 10 & 0.0062&144& 0.80& 10 & G341/HG31 \\
	NGC 5129& 13 24 10.0 & +13 58 42.3& 38 & 0.0233& 304&   0.65  & 12 &
	 GH117/N117  \\
	NGC 5171 & 13 29 24.3 & +11 44 11.9& 15 & 0.0232& 477 &0.00 & 5 & - \\ 
	NGC 5218& 13 32 42.3& +62 34 41.5& 3 & 0.0103&132 &0.66 &10 & G354/GH122/N120 \\
	IC 4296  & 13 36 41.5& -34 00 32.4 &29&0.0125&228&0.69 & 13 &- \\ 
	HCG 67   & 13 49 12.3 & -07 12 34.3& 14 & 0.0248& 351& 0.50 & 4,5 & H67  \\
	NGC 5322 & 13 49 14.7& +60 11 27.4 & 8 & 0.0065& 196 & 0.37 & 1 & G360/N123 \\
	HCG 68  & 13 53 27.1 & +40 16 55.9& 5 & 0.0080& 155& 0.20 &2,3 & G363/H68/N124\\
	NGC 5374& 13 58 00.1 & +06 15 25.0& 3 & 0.0143& 9  & 0.33 & 1 & G368/GH127/N129  \\
	NGC 5775& 14 53 24.9 & +03 29 47.5& 5 & 0.0051& 88 & 0.80 & 1 &G387/GH148/N143  \\
	NGC 5846 & 15 06 29.4& +01 36 12.3& 20 & 0.0063& 368&  0.73 & 9 &G393/GH150/HG50/N146  \\
	NGC 5866 & 15 16 23.7& +56 25 01.9& 4 & 0.0022& 74 & 0.75  & 1 &G396/GH152/HG78/N147 \\
	NGC 5929 & 15 26 18.1& +41 05 34.2& 3 & 0.0086& 66 & 1.00 & 1 & G399/GH153/N150 \\
	NGC 5970& 15 36 16.2& +12 02 07.7& 3 & 0.0064& 81 & 1.00 & 1 &G401/GH157/N154  \\
	NGC 6052& 15 59 31.1& +20 32 33.0
	 & 11 & 0.0154&183 &0.82 & 10 & G403/GH161/N158\\
	NGC 6109 & 16 17 26.9& +34 55 23.5& 53 & 0.0312& 563&0.53 & 15 & - \\
	NGC 6251 & 16 32 32.3& +82 32 15.0& 2 & 0.0221& - & 0.00& 5& - \\
	HCG 84   & 16 44 08.1& +77 50 09.3& 5 &0.0556 &204 & 0.20&2,3 & H84 \\
	ARP 330  & 16 49 11.8& +53 25 12.0& 8 & 0.0298& 369 & 0.20 & 5& - \\
	NGC 6269 & 16 57 54.4 & +27 52 17.8& 46& 0.0353 &586 &0.35& 8,5& -  \\
NGC 6329 & 17 14 14.4& +43 41 01.2& 13 & 0.0274 & 419& 0.83& 5&-  \\
NGC 6338 & 17 15 22.9& +57 24 38.1& 11& 0.0284 &538 & 0.38& 5& - \\ 
	HCG 90& 22 01 59.5& -31 57 50.4&16& 0.0085&193&0.75 &9 &H90/G450/MDL59 \\
UGC 12064 & 22 31 21.3& +39 21 21.7& 9& 0.0166 &399 & 0.75 & 5& - \\
HCG 92   & 22 35 59.0& +33 57 11.9& 4 & 0.0215 & 389& 0.75 & 2,3 & G459/H92  \\
NGC 7358& 22 46 44.2& -65 04 10.5& 3 & 0.0098& 113 & 0.00 & 11 & G462/MDL5 \\
IC 1459 & 22 57 10.4& -36 27 37.4& 5 & 0.0057 &80& 0.80&10 & G466/HG15 \\   
NGC 7448& 23 01 04.9 &+15 52 07.9& 6 & 0.0068 & 116& 0.33 & 1 & G469/N160 \\
HCG 93   & 23 15 24.2& +18 58 59.2& 4 & 0.0168 & 209& 0.50 & 2,3 & H93 \\
NGC 7582& 23 18 15.3& -42 31 08.2& 8 &  0.0054 & 38&1.00 & 9&G472/HG12 \\
NGC 7619& 23 20 14.8& +08 12 27.7& 7 & 0.0116 & 253& 0.14 & 1 & G473/GH166/N164 \\
HCG 96   & 23 27 58.3& +08 46 26.6& 4 & 0.0292 & 132& 0.75 & 2,3 & H96  \\
HCG 97   & 23 47 23.2& -02 18 07.4& 14 & 0.0221 & 425& 0.50 & 4,5 & H97  \\
NGC 7777& 23 53 09.1& +28 15 44.3& 4 & 0.0229 & 116 & 0.25 & 1 & N170 \\ 
\enddata

\tablecomments{$^1$ G=Garcia (1993), GH=Geller \& Huchra (1983),
H=Hickson (1982), HG=Huchra \& Geller (1982), MDL=Maia, Da Costa \& Latham (1989
), N=Noltenius (1993)}

\tablecomments{references: 1) Nolthenius 1993, 2) Hickson 
et al. 1992, 3) Hickson, Kindl and Auman 1989, 4) Ribeiro et al. 1998
 5) NASA Extragalactic Database (NED), 6) Geller and Huchra 1983,
7) Werner et al. 1999, 8) Zimer, Zabludoff and Mulchaey (2002), 
9) Zabludoff and Mulchaey (1998),
10) Garcia 1993, 11) Maia, Da Costa and Latham 1989, 12) Zabludoff and Mulchaey (2000),
13) Willmer et al. 1999, 14) Willmer et al. 1991,
15) Mahdavi et al. 1999.}

\end{deluxetable}

%% file: table2.tex
\begin{deluxetable}{rcccccccr}
\tablecolumns{10}
\tablewidth{0pc}
\tablecaption{X-ray Properties}
\tablehead{
\colhead{Group}& \colhead{Exposure} & \colhead{Detection}& \colhead{N$_{\rm H}$}  & \colhead{R$_{\rm X}$} &
\colhead{T} & \colhead{Abundance} & \colhead{D} & \colhead{Log L$_{\rm x}$} \\
  & \colhead{(sec)} & & \colhead{(cm$^{-2}$)} & \colhead{(arcmin/kpc)} & \colhead{(keV)} &
\colhead{(solar)} & \colhead{(Mpc)} & \colhead{(erg/s)} \\
}\startdata
NGC 7805& 13,995& UL  & 5.6& - & - & - & 46.4 & $<$40.31 \\
NGC 43  & 7,822 &UL  & 5.5& - & - & - & 45.2 & $<$40.85 \\
HCG 2   & 18,633 & UL  & 4.3& - &  - & - & 40.0 & $<$40.85 \\
HCG 4   & 8,852  &UL  & 1.6& - & - & - & 71.7 & $<$41.11 \\
HCG 3   & 7,156  &UL  & 3.6& - & - & - & 78.6 & $<$41.54 \\
IC 1559 & 6,841  &UL  & 3.4& - & - & - & 44.5 & $<$40.90 \\
NGC 315 & 25,939&DE/AGN &5.9& 5/66 &0.80$^{+0.18}_{-0.14}$ &0.06$^{+0.09}_{-0.04}$ & 46.6
& 41.57$^{+0.10}_{-0.08}$ \\
        && &    & & 0.81$^{+0.19}_{-0.12}$ & (0.3) & &  \\

NGC 326 & 19,680 &DE &5.8 & 19/734 & 1.41$^{+0.72}_{-0.37}$ & 0.32$^{+0.55}_{-0.23}$ &144.8 
& 42.97$^{+0.12}_{-0.12}$ \\
        && &    & & 1.41$^{+0.47}_{-0.24}$ & (0.3) & &  \\
NGC 383 &28,220&DE & 5.4& 30/442 & 1.51$^{+0.21}_{-0.11}$ & 0.12$^{+0.10}_{-0.07}$ & 49.5 
& 42.72$^{+0.05}_{-0.04}$ \\
        &&   &  &&  3.37$^{+0.59}_{-0.55}$ & (0.3) &  & \\
NGC 491 &10,450& UL  & 2.3& -& - & - & 35.7 & $<$40.70 \\
NGC 507 &17,878& DE & 5.2& 33/455 & 1.25$^{+0.08}_{-0.07}$ & 0.22$^{+0.06}_{-0.05}$ & 48.7
& 42.95$^{+0.03}_{-0.03}$ \\
        &&    & && 1.30$^{+0.05}_{-0.05}$ & (0.3) & & \\
NGC 524 &10,560& DE & 4.5&9/57 & 0.47$^{+0.24}_{-0.19}$ & (0.3) & 21.9 & 40.53$^{+0.37}_{-0.47}$ \\
NGC 533 &11,779&DE & 3.1&18/263 & 1.03$^{+0.06}_{-0.08}$ & 0.59$^{+0.86}_{-0.26}$&  51.8
& 42.37$^{+0.24}_{-0.13}$ \\
        &&   &  && 1.00$^{+0.07}_{-0.09}$ & (0.3)     & &  \\
HCG 10  &14,131& UL & 4.9&- & - & - & 46.0 & $<$40.67 \\
HCG 12  &8,923& DE &3.6& 5/197 & 0.67$^{+0.61}_{-0.31}$ & (0.3) & 147.5 & 41.88$^{+0.20}_{-0.45}$ \\
NGC 584 &3,706& UL &3.8&- & - & - & 16.9 & $<$40.51 \\
NGC 664 &7,531& UL &3.4 &- & - & - & 52.3 & $<$40.98 \\ 
NGC 720 &22,487& DE & 1.5 & 9/39& 0.51$^{+0.07}_{-0.05}$ & 0.09$^{+0.06}_{-0.04}$ & 15.1 
& 40.86$^{+0.09}_{-0.07}$ \\
        &&    & && 0.53$^{+0.06}_{-0.05}$ & (0.3) & & \\
NGC 741 &12,865& DE & 4.5&16/240 & 1.06$^{+0.13}_{-0.12}$ & 0.22$^{+0.22}_{-0.11}$ & 53.4 &
 42.14$^{+0.11}_{-0.09}$\\
        & &   & &  &  1.08$^{+0.13}_{-0.09}$ & (0.3) & &  \\
HCG 15  & 13,228&DE? &3.2& 9/168 & 1.26$^{+0.53}_{-0.33}$ & (0.3) & 66.8 & 41.72$^{+0.18}_{-0.22}$ \\
HCG 16  &14,000& DE &2.2&11/118 & 0.37$^{+0.40}_{-0.14}$ & (0.3) & 37.1 & 40.74$^{+0.38}_{-0.49}$ \\
UGC 1651&13,424& DE &6.1&11/328 & 1.43* & (0.3) & 109.3 & 42.15$^{+0.16}_{-0.22}$ \\
HCG 18  &2,942& UL  & 8.6 & - & - & &39.1 &$<$41.04 \\
NGC 1044&16,184& DE & 8.8 &8/141&  0.87$^{+0.61}_{-0.29}$ & 0.06$^{+0.62}_{-0.06}$ & 62.8 & 41.69$^{+0.42}_{-0.13}$ \\
        & &  & & & 0.99$^{+0.74}_{-0.33}$ & (0.3) & & \\
IC 1860 & 9,180 & DE & 2.1&18/345 & 1.14$^{+0.20}_{-0.11}$ & 0.17$^{+0.14}_{-0.07}$ & 
68.8 &42.75$^{+0.06}_{-0.06}$ \\
        & &   & && 1.26$^{+0.14}_{-0.17}$ & (0.3) & & \\
HCG 22  &12,298& DE? & 4.2 & -& - & -&25.4 & $<$40.37  \\
IC 1880 &4,122& DE & 6.5& 7/196& 1.13$^{+0.20}_{-0.16}$& (0.3) & 102.0 & 42.34$^{+0.22}_{-0.24}$ \\
HCG 23  &4,122& UL & 6.5 & -& - & - & 44.3  & $<$40.96 \\
HCG 26  &5,011& UL & 5.4 & -& - & - & 95.4 & $<$41.26\\
NGC 1332&25,051& UL & 2.2 & -&- & - & 14.2 & $<$40.04\\
UGC 2755&16,826& DE & 14.3& 7/134&  0.66$^{+0.22}_{-0.21}$ & (0.3) & 68.6 & 41.60$^{+0.33}_{-0.41}$ \\
NGC 1399&26,954& DE & 1.3 &40/115 & 1.38$^{+0.03}_{-0.03}$ & 0.28$^{+0.04}_{-0.04}$ & 9.9 &41.94$^{+0.04}_{-0.03}$ \\
        &      &    &    &  & 1.39$^{+0.02}_{-0.02}$ & (0.3)  &      &      \\
NGC 1407&18,222& DE & 5.4 & 15/69& 0.95$^{+0.09}_{-0.07}$ & 0.14$^{+0.10}_{-0.05}$ &  15.9
& 41.23$^{+0.07}_{-0.05}$ \\
        & & &   &   & 1.02$^{+0.06}_{-0.07}$ & (0.3) & & \\
NGC 1587& 8,080&DE & 6.5& 6/62 & 0.90$^{+0.81}_{-0.46}$& (0.3) & 36.2 & 40.92$^{+0.13}_{-0.41}$ \\
HCG 31  & 1,584&UL & 6.0&- & - & - & 39.5 & $<$40.97\\
HCG 33  & 4,081&DE?& 22.0& -& - & - & 79.3 &$<$41.45 \\
NGC 1961& 10,703&UL & 8.3 & -& - & - & 41.8 & $<$40.84 \\
NGC 2300& 17,232&DE &5.1& 25/135 & 0.87$^{+0.12}_{-0.09}$ & 0.04$^{+0.03}_{-0.02}$ & 20.5 
& 41.62$^{+0.04}_{-0.04}$ \\
        &&   &   &   & 1.05$^{+0.06}_{-0.07}$ & (0.3) & & \\
NGC 2484&15,504& DE/AGN & 5.1& 7/249& 0.86$^{+0.63}_{-0.38}$ & 0.04$^{+0.31}_{-0.04}$ & 130.1 &42.41$^{+0.21}_{-0.15}$ \\
        &&  &   &  &  1.04$^{+0.92}_{-0.34}$ & (0.3) & & \\ 
NGC 2563&21,155& DE& 4.3& 18/258 & 1.09$^{+0.10}_{-0.10}$ & 0.18$^{+0.13}_{-0.07}$ & 50.6
& 42.16$^{+0.07}_{-0.06}$ \\
        &&  &   &   & 1.11$^{+0.15}_{-0.07}$ & (0.3) & & \\
HCG 35  &15,262& DE? & 2.8& - & - & - & 171.1 &$<$41.84 \\
NGC 2769&8,773& UL & 1.6 & -& - & - & 52.1 & $<$40.82 \\
HCG 37&4,884& DE &1.6& 8/157 &  0.65$^{+0.29}_{-0.32}$ & (0.3) & 70.6 & 41.63$^{+0.31}_{-0.46}$ \\
NGC 2805& 5,931& UL &3.9 &- & - & - & 17.6 & $<$40.46 \\
HCG 38  & 7,692&UL & 3.5 &- & - & - & 92.8 & $<$41.40 \\
HCG 40  & 9,459&UL& 3.5 & -& - & - & 70.9 & $<$41.05 \\
HCG 42  & 11,799& DE & 4.8& 9/108 & 0.73$^{+0.08}_{-0.06}$ & (0.3) & 42.3& 41.59$^{+0.32}_{-0.34}$ \\
HCG 44  & 4,543&UL& 2.1 & -&- & - & 17.0 & $<$40.70 \\
MKW 2   & 9,017 &DE& 4.4& 9/286 & 1.31$^{+0.89}_{-0.28}$& (0.3) & 116.3 & 42.32$^{+0.13}_{-0.21}$ \\
HCG 48  & 17,972&DE?& 5.1&6/54& 2.56$^{+2.4}_{-0.75}$ & (0.3) & 31.9& 41.39$^{+0.19}_{-0.13}$ \\
CGCG154-041&20,581& DE& 1.9& 8/242 &1.05$^{+0.32}_{-0.22}$ & (0.3) & 110.7 & 41.99$^{+0.20}_{-0.28}$ \\
NGC 3396&15,493& UL& 2.0 & -&- & - & 18.8 &$<$40.49$^1$  \\
NGC 3557&19,219& DE & 7.4 &8/73&  0.25$^{+0.08}_{-0.04}$ & (0.3) & 31.9 & 41.12$^{+0.32}_{-0.58}$ \\
NGC 3607&23,143& DE &1.6&10/42 & 0.40$^{+0.06}_{-0.05}$ & (0.3) & 14.5 & 40.53$^{+0.38}_{-0.42}$ \\
HCG 50  &1,175& UL & 0.8 &-& - & - & 461.9 & $<$42.55 \\
NGC 3647&13,691& DE & 4.3&18/759 & 1.46$^{+0.62}_{-0.33}$ & 0.02$^{+0.09}_{-0.02}$& 157.9 & 43.37$^{+0.24}_{-0.11}$\\
        &  &&  & & 2.21$^{+0.98}_{-0.49}$& (0.3) & & \\
NGC 3665&6,613& DE & 2.1&6/40 & 0.33$^{+0.14}_{-0.07}$ & (0.3) & 23.4 & 40.63$^{+0.37}_{-0.47}$\\
HCG 57  &10,640& DE?& 2.0&8/203& 1.14$^{+1.93}_{-0.31}$ & (0.3) & 96.6 & 41.92$^{+0.14}_{-0.27}$ \\
HCG 58  & 11,321& UL & 3.2& -& - & - & 66.4 & $<$40.88$^1$ \\ 
NGC 3923&36,697& DE &6.2&8/40 & 0.52$^{+0.12}_{-0.13}$ & (0.3) & 17.3 & 40.29$^{+0.37}_{-0.42}$ \\ 
NGC 4065&15,422& DE &2.4&17/351 & 1.20$^{+0.19}_{-0.13}$ & (0.3) & 74.1 & 42.40$^{+0.21}_{-0.22}$ \\
NGC 4073&9,697& DE &1.9 &22/404& 1.75$^{+0.24}_{-0.15}$ & 1.14$^{+0.72}_{-0.43}$ & 64.9 
& 43.04$^{+0.09}_{-0.08}$ \\
        &&   &  &   & 1.44$^{+0.12}_{-0.06}$ & (0.3) & & \\
NGC 4104&13,891& DE &1.7&17/427 & 1.46$^{+0.31}_{-0.16}$ & 0.29$^{+0.29}_{-0.14}$ & 90.7 &
 42.77$^{+0.08}_{-0.06}$ \\
        &&    & &  &  1.48$^{+0.24}_{-0.14}$ & (0.3) & & \\
NGC 4125&3,900& DE &1.9& 6/28 & 0.44$^{+0.18}_{-0.11}$ & (0.3) & 16.1 & 40.54$^{+0.38}_{-0.43}$ \\
NGC 4168& 13,824& UL&2.6&-& - & - & 26.6 & $<$40.56 \\
NGC 4261&19,163& DE & 1.5&27/187 & 0.91$^{+0.17}_{-0.11}$ &  0.03$^{+0.04}_{-0.03}$& 24.0
& 41.89$^{+0.07}_{-0.05}$ \\
        &&    & &   & 1.29$^{+0.11}_{-0.11}$ & (0.3) & & \\
SHK 202 &11,563& DE & 1.9& 15/340& 0.77$^{+0.29}_{-0.20}$ & (0.3) & 81.6 & 42.01$^{+0.27}_{-0.37}$ \\
NGC 4278&3,228& UL& 1.8&-&-  & - & 12.5 &$<$40.83$^1$  \\
NGC 4357&3,139& UL & 1.3& -& - & - & 45.0 & $<$41.37 \\
NGC 4325&4,984& DE & 2.2&9/203 & 0.81$^{+0.04}_{-0.04}$ & 0.45$^{+0.44}_{-0.17}$& 81.1
& 42.80$^{+0.19}_{-0.11}$ \\
        &&     & &  &  0.81$^{+0.04}_{-0.05}$ & (0.3) & & \\
NGC 4291&11,502& DE & 2.8&14/84 & 0.49$^{+0.25}_{-0.20}$ & (0.3) & 20.9 & 40.74$^{+0.37}_{-0.46}$ \\
NGC 4615&4,845& UL & 1.2& -& - & - & 50.9 & $<$41.00 \\
NGC 4636&11,737& DE & 1.8&22/105& 0.74$^{+0.04}_{-0.03}$ & 0.41$^{+0.40}_{-0.12}$ & 16.6 
& 41.88$^{+0.20}_{-0.08}$ \\
        &&     & &   & 0.73$^{+0.03}_{-0.03}$ & (0.3) & & \\
HCG 62  &17,711& DE &3.0& 29/389 & 0.96$^{+0.08}_{-0.08}$ & 0.09$^{+0.06}_{-0.04}$ & 47.4
& 42.69$^{+0.05}_{-0.05}$ \\
        &&    & &   & 1.08$^{+0.05}_{-0.05}$ & (0.3)& & \\
NGC 5044&26,097& DE &5.0&36/287& 1.02$^{+0.02}_{-0.02}$ & 0.20$^{+0.03}_{-0.03}$& 27.9
& 42.81$^{+0.02}_{-0.02}$ \\
        &&    & &   & 1.05$^{+0.01}_{-0.02}$ & (0.3)& & \\
NGC 5101&5,278& UL & 5.7&- & - & - & 21.7 &$<$40.77 \\
NGC 5129&5,568& DE & 1.8 &9/187& 0.81$^{+0.12}_{-0.10}$ & (0.3)& 74.1 & 42.16$^{+0.31}_{-0.33}$ \\
NGC 5171& 4,616& DE &1.9 &14/288& 1.25$^{+0.40}_{-0.25}$ & (0.3)& 73.8 & 42.35$^{+0.17}_{-0.23}$ \\
NGC 5218&5,653& UL & 2.0& - &- & - & 32.1 & $<$40.93 \\
IC 4296 &3,774& DE &4.1 &15/173& 0.85$^{+0.33}_{-0.24}$ & (0.3)& 40.6 & 41.85$^{+0.24}_{-0.35}$ \\
HCG 67  &15,093& DE &2.5 &10/219& 0.78$^{+0.32}_{-0.22}$ & (0.3)& 78.8 & 41.65$^{+0.26}_{-0.37}$ \\
NGC 5322&26,585& DE &1.8 &5/29& 0.33$^{+0.19}_{-0.07}$& (0.3)& 20.6 & 40.10$^{+0.38}_{-0.47}$ \\
HCG 68  &11,780& DE &1.0 &10/75& 0.59$^{+0.10}_{-0.11}$& (0.3)& 26.0 & 40.99$^{+0.36}_{-0.39}$ \\
NGC 5374&5,556& UL &2.1 & -&- & - & 46.1 &$<$41.05 \\
NGC 5775&5,913& UL &3.5 & -&- & - & 17.4 &$<$40.62 \\
NGC 5846&6,979& DE& 4.2 & 15/90& 0.70$^{+0.05}_{-0.04}$& 0.31$^{+0.28}_{-0.12}$ & 20.9 
& 41.71$^{+0.17}_{-0.11}$ \\
        &&   &  &   & 0.70$^{+0.04}_{-0.04}$& (0.3)& & \\
NGC 5866&16,332& UL &1.5&-& - & - & 7.2 & $<$39.48$^1$ \\
NGC 5929&10,347& UL &2.0& -&- & - & 26.9 &$<$40.50 \\
NGC 5970&6,866& UL &3.5& - &-& - & 20.6 &$<$41.37 \\
NGC 6052&4,728& UL &4.0& -&- & - & 47.6 &$<$41.13 \\
NGC 6109&6,020& DE &1.4 &25/662& 2.22$^{+1.20}_{-0.58}$& 0.13$^{+0.43}_{-0.13}$ & 96.2
& 43.21$^{+0.14}_{-0.10}$ \\
        &&    & &   & 2.43$^{+1.28}_{-0.53}$& (0.3) & & \\
NGC 6251&13,699& DE/AGN&  5.5&10/187 & 0.87$^{+0.25}_{-0.17}$& (0.3) & 67.0 & 41.82$^{+0.26}_{-0.33}$ \\
HCG 84  &35,363& UL & 4.0 & -&- & - & 173.3 &$<$41.93 \\
ARP 330 &6,794& DE & 2.7 &10/252& 1.29$^{+0.14}_{-0.20}$ & (0.3) & 91.2 & 42.57$^{+0.20}_{-0.21}$ \\
NGC 6269&10,345& DE &4.7 &23/703& 1.83$^{+0.62}_{-0.27}$ & 0.27$^{+0.33}_{-0.16}$ & 108.7 
& 43.20$^{+0.09}_{-0.07}$\\ 
        &&    & &   & 1.92$^{+0.44}_{-0.26}$ & (0.3) &  & \\
NGC 6329&17,995& DE &2.1 &16/370& 0.99$^{+0.10}_{-0.12}$ & 0.16$^{+0.15}_{-0.08}$ & 83.5 &
 42.45$^{+0.09}_{-0.08}$  \\
        &&    & &   & 1.03$^{+0.09}_{-0.10}$ & (0.3) & & \\
NGC 6338&4,006& DE &2.6 &15/359& 1.82$^{+0.90}_{-0.40}$ & 0.19$^{+0.32}_{-0.15}$ & 86.6
& 43.25$^{+0.11}_{-0.09}$ \\
        &&    & &   & 2.02$^{+0.83}_{-0.37}$ & (0.3) & & \\
HCG 90&17,146& DE & 1.6 &10/66& 0.51$^{+0.15}_{-0.14}$ & (0.3) & 22.9 & 40.60$^{+0.37}_{-0.42}$ \\
UGC 12064&9,068& DE& 
11.8&15/200 & 1.03$^{+0.14}_{-0.14}$ & (0.3) & 47.2 & 42.17$^{+0.26}_{-0.27}$ \\
HCG 92 &18,695& DE&  8.3& 5/92 &  1.05$^{+0.89}_{-0.30}$ & (0.3) & 65.2 & 41.31$^{+0.12}_{-0.30}$ \\
NGC 7358&11,064& UL & 2.5&- & - & - & 28.4 &$<$40.59 \\
IC 1459 &31,139& DE &1.2& 12/50 & 0.63$^{+0.06}_{-0.06}$ & (0.3) & 14.4 & 40.52$^{+0.36}_{-0.36}$ \\
NGC 7448&6,762& UL&6.1 & - & -&- & 17.8 &$<$40.40 \\ 
HCG 93&15,022& UL & 4.7 &-& - & - & 47.3 &$<$40.65 \\
NGC 7582&7,028& UL & 1.9 &-& - & - & 13.7 &$<$40.74$^1$ \\
NGC 7619&17,943& DE &5.0&23/96 & 0.98$^{+0.07}_{-0.07}$ & 0.12$^{+0.06}_{-0.04}$ & 31.3
& 42.05$^{+0.05}_{-0.04}$\\ 
        &&    & &   & 1.05$^{+0.05}_{-0.05}$ & (0.3) & & \\
HCG 96  &2,480& UL & 5.2&- & - & - & 85.6 &$<$41.64 \\
HCG 97  &12,092& DE & 3.7&11/196 & 0.97$^{+0.14}_{-0.12}$ & 0.16$^{+0.10}_{-0.09}$ & 63.6
& 42.06$^{+0.07}_{-0.10}$ \\
        &&     & & & 1.02$^{+0.11}_{-0.12}$ & (0.3) & & \\
NGC 7777&16,924& DE & 5.0&7/130 & 0.63$^{+0.52}_{-0.20}$& (0.3) & 66.3 & 41.28$^{+0.25}_{-0.42}$ \\
\enddata
\tablecomments{
$^1$ Extraction radius for the X-ray luminosity is less than 200 kpc: NGC 3396 (194 kpc), HCG 58 (83 kpc), NGC 4278 (148 kpc), NGC 5866 (76 kpc), NGC 7582 (120 kpc).}
\end{deluxetable}


%% file: table3.tex
\begin{deluxetable}{rccccc}
\tablecolumns{6}
\tablewidth{0pc}
\tablecaption{Spatial Fits}
\tablehead{
\multicolumn{1}{r}{Group} & \multicolumn{3}{c}{Extended Component} &
\multicolumn{2}{c}{Central Component} \\ 
\colhead{} & \colhead{R$_{\rm core}$} & \colhead{$\beta$} 
& \colhead{ellipticity} & \colhead{R$_{\rm core}$}
 & \colhead{$\beta$} \\
\colhead{} & \colhead{(arcminute)} & \colhead{} & \colhead{} & \colhead{(arcminute)} & 
\colhead{} \\ }
 \startdata
NGC 315 & 0.23$\pm{0.02}$ & 0.77$\pm{0.02}$ & 0.00$^{+0.05}_{-0.00}$  & - & - \\
NGC 326 & 2.00$\pm{0.15}$ & 0.36$\pm{0.01}$ & 0.05$^{+0.11}_{-0.05}$ & - &
 - \\
        & 2.22$\pm{0.19}$ & 0.35$\pm{0.01}$ & 0.09$^{+0.11}_{-0.09}$ & 
0.11$\pm{0.02}$ & 0.50$\pm{0.04}$ \\
NGC383  & $<$0.1    & 0.32$\pm{0.01}$ & 0.13$\pm{0.04}$ & - & - 
  \\
        & $<$0.1 & 0.42$\pm{0.01}$ & 0.12$\pm{0.05}$ & 0.22$\pm{0.01}$ & (1.0) \\
NGC 507 & 1.28$\pm{0.03}$ & 0.46$\pm{0.01}$ & 0.09$\pm{0.03}$ & - & - \\
        & 1.27$\pm{0.03}$ & 0.42$\pm{0.01}$ & 0.09$\pm{0.03}$ &
$<$0.1 & (1.0) \\
NGC 524 & $<$0.1    & 0.41$\pm{0.02}$ & 0.15$^{+0.18}_{-0.15}$ & - 
& - \\
NGC 533 & $<$0.1   & 0.46$\pm{0.01}$ & 0.10$\pm{0.05}$ 
& - 
& - \\ 
        &  0.42$\pm{0.02}$ & 0.95$\pm{0.04}$ &0.21$\pm{0.05}$ &
$<$0.1 & 0.36$\pm{0.01}$ \\
HCG 12 & $<$0.1 & 0.34$\pm{0.02}$ & 0.53$^{+0.25}_{-0.34}$ & - & -
\\
NGC 720& $<$0.1 & 0.41$\pm{0.01}$ & 0.15$\pm{0.05}$ &-
& - \\
       & 0.44$\pm{0.02}$ & 0.65$\pm{0.01}$ & 0.14$\pm{0.05}$ &
$<$0.1 & 0.31$\pm{0.01}$  \\
NGC 741 & $<$0.1    & 0.43$\pm{0.01}$ & 0.00$^{+0.08}_{-0.00}$ & - & -
\\
        & $<$0.1 & 0.43$\pm{0.01}$ & 0.00$^{+0.08}_{-0.00}$ & $<$0.1 & (1.0) \\
HCG 15 & $<$0.1 & 0.34$\pm{0.01}$ & 0.18$^{+0.25}_{-0.18}$ & - & - 
\\
HCG 16 & $<$0.1 & 0.35$\pm{0.03}$ & (0.00) & - & - \\
UGC 1651 & $<$0.1 & 0.33$\pm{0.01}$ & 0.21$^{+0.16}_{-0.18}$ &
 - & - 
\\
         & $<$0.1 & 0.38$\pm{0.01}$ & 0.19$\pm{0.19}$ & 0.14$\pm{0.02}$ & 0.77$^{+0.11}_{-0.07}$ \\
NGC 1044 & $<$0.1 & 0.38$\pm{0.02}$ & 0.26$^{+0.22}_{-0.20}$ & - & - 
\\
IC 1860  & 0.22$\pm{0.01}$ & 0.46$\pm{0.01}$ & 0.00$^{+0.04}_{-0.00}$ & - 
& - 
\\
         & 0.32$\pm{0.01}$ & 0.47$\pm{0.02}$ & 0.00$^{+0.04}_{-0.00}$ & 0.30$\pm{0.01}$ & (1.0) \\
IC 1880  & $<$0.1 & (1.00) & (0.00) & - & - \\
UGC 2755 & $<$0.1 & 0.36$\pm{0.01}$ & 0.11$^{+0.13}_{-0.11}$ & - & -
\\
         & $<$0.1 & 0.48$\pm{0.02}$ & 0.22$^{+0.13}_{-0.13}$ & 
$<$0.1 & 0.31$\pm{0.01}$ \\
NGC 1399 & $<$0.1 & 0.39$\pm{0.01}$ & 0.03$\pm{0.01}$ & - & -
\\
         & $<$0.1 & 0.38$\pm{0.01}$ & 0.02$\pm{0.01}$ & 0.32$\pm{0.01}$ & (1.0) \\
NGC 1407 & 0.20$\pm{0.02}$ & 0.34$\pm{0.01}$ & 0.23$\pm{0.12}$ & - & - \\
         & 0.60$\pm{0.05}$ & 0.37$\pm{0.01}$ & 0.22$\pm{0.11}$ & $<$0.1 & (1.0) \\
NGC 1587 & $<$0.1 & 0.42$\pm{0.02}$ & (0.00) & - & - \\
NGC 2300 & $<$0.1 & (1.0) & 0.39$^{+0.09}_{-0.06}$ &
- & - \\
NGC 2484 & $<$0.1 & 0.53$\pm{0.01}$ & 0.11$\pm{0.09}$ & - & -
\\
NGC 2563 & 1.28$\pm{0.11}$ & 0.33$\pm{0.01}$ & 0.15$^{+0.11}_{-0.12}$ &
- & - \\
         & 1.29$\pm{0.10}$ & 0.42$\pm{0.01}$ & 0.14$\pm{0.11}$ &
$<$0.1 & (1.0) \\
HCG 37 & $<$0.1 & 0.31$\pm{0.03}$ & (0.00) & - & - \\
HCG 42 & 0.13$\pm{0.03}$& 0.35$\pm{0.01}$& (0.00)& - & - \\
       & 1.75$\pm{0.70}$ & 0.41$\pm{0.02}$ & (0.00) & 0.84$\pm{0.45}$ & 
(1.00) \\
MKW 2 & $<$0.1 & 0.34$\pm{0.02}$ & 0.00$^{+0.37}_{-0.00}$ & - & - \\
      & $<$0.1 & 0.33$\pm{0.02}$ & 0.17$^{+0.36}_{-0.17}$ & $<$0.1 & (1.0) \\
HCG 48 & 0.85$\pm{0.05}$ & 0.56$\pm{0.01}$ & 0.00$^{+0.06}_{-0.00}$ 
& - & - \\
CGCG154-041& $<$0.1 & 0.38$\pm{0.01}$ & 0.14$\pm{0.11}$ &
- & - \\
  & $<$0.1 & 0.36$\pm{0.01}$ & 0.15$\pm{0.14}$ &  0.23$\pm{0.05}$ & (1.0) \\
NGC 3557 & $<$0.1 & 0.44$\pm{0.01}$ & 0.00$^{+0.13}_{-0.00}$ & - & - \\
NGC 3607 & $<$0.1 & 0.43$\pm{0.01}$ & 0.20$\pm{0.11}$ &
- & - \\
         & $<$0.1 & 0.42$\pm{0.01}$ & (0.00)  & $<$0.1 & 0.50$\pm{0.02}$ \\
NGC 3647 & 2.51$^{+0.24}_{-0.14}$ & 0.40$\pm{0.01}$  & 0.00$^{+0.10}_{-0.00}$ &
- & - \\
         & 4.34$\pm{0.35}$ & 0.46$\pm{0.02}$ & 0.10$^{+0.10}_{-0.11}$ &
$<$0.1 & (1.00) \\
NGC 3665 & $<$0.1 & 0.42$\pm{0.03}$ & (0.00) & - & - \\
HCG 57   & 0.25$\pm{0.10}$ & 0.34$\pm{0.02}$ & (0.00) & - & - \\
NGC 3923 & $<$0.1 & 0.56$\pm{0.01}$ & 0.00$^{+0.03}_{-0.00}$ & - & - \\
         & 0.34$\pm{0.01}$ & 0.89$\pm{0.02}$ & (0.00) & $<$0.1 & 0.35$\pm{0.01}$ \\
NGC 4065 & 3.75$^{+0.59}_{-0.16}$ & 0.42$\pm{0.01}$ & (0.00) & - & - \\
         & 2.52$^{+0.56}_{-0.46}$ & 0.46$\pm{0.01}$ & 0.45$^{+0.15}_{-0.20}$ &
$<$0.1 & 0.31$\pm{0.01}$ \\
NGC 4073 & $<$0.1 & 0.44$\pm{0.01}$ & 0.15$\pm{0.03}$ & - 
& - \\
         & $<$0.1 & 0.43$\pm{0.01}$ & 0.19$\pm{0.04}$ & 0.61$\pm{0.04}$ & (1.00) \\
NGC 4104 & 0.95$\pm{0.07}$ & 0.38$\pm{0.01}$ & 0.00$^{+0.09}_{-0.00}$ & 
- & - \\
         & 0.95$\pm{0.08}$ & 0.42$\pm{0.01}$ &0.00$^{+0.11}_{-0.00}$ &
$<$0.1 & 0.36$\pm{0.01}$ \\
NGC 4125 & $<$0.1 & 0.47$\pm{0.03}$ & (0.00) & - & - \\
NGC 4261 & $<$0.1 & (1.00) & (0.00) & - & - \\
NGC 4325 & 0.40$\pm{0.01}$ & 0.65$\pm{0.01}$ & 0.00$^{+0.03}_{-0.00}$ & - 
& - \\
         & 0.90$\pm{0.03}$ & 1.13$\pm{0.06}$ & (0.00) & $<$0.1 & 0.43$\pm{0.01}$ \\ 
NGC 4636 & 0.16$\pm{0.01}$ & 0.43$\pm{0.01}$ & 0.00$^{+0.01}_{-0.00}$ &
- & - \\
         & 0.77$\pm{0.01}$ & 0.76$\pm{0.01}$ & (0.00) & $<$0.1 & 0.33$\pm{0.01}$ \\
HCG 62 & $<$0.1 & 1.76$^{+0.23}_{-0.17}$ & 0.00$^{+0.16}_{-0.00}$ & 
- & - \\
       & $<$0.1 & 0.35$\pm{0.01}$ & 0.18$\pm{0.05}$ & 1.34$\pm{0.02}$ & 2.17$^{+0.05}_{-0.08}$ \\
NGC 5044 & 1.28$\pm{0.01}$ & 0.52$\pm{0.01}$ & 0.00$^{+0.03}_{-0.00}$ &
- & - \\
         & 0.76$\pm{0.01}$ & 0.43$\pm{0.01}$ & 0.14$\pm{0.01}$ & 3.88$\pm{0.03}$ & 1.14$\pm{0.01}$ \\
NGC 5129 & $<$0.1 & 0.39$\pm{0.01}$ 
& 0.00$^{+0.17}_{-0.00}$ & - & - \\         
NGC 5171 & 0.28$\pm{0.11}$ & 0.32$\pm{0.02}$& 0.29$^{+0.32}_{-0.29}$& - & - \\
IC 4296  & 3.47$^{+1.65}_{-1.24}$ & 0.35$\pm{0.03}$ & (0.00) & - & - \\
HCG 67   & $<$0.1 & 0.36$\pm{0.01}$ & 0.00$^{+0.18}_{-0.00}$ & - & - \\
NGC 5322 & $<$0.1 & 0.38$\pm{0.01}$ & 0.41$\pm{0.13}$ & - & - \\
HCG 68   & $<$0.1 & 0.43$\pm{0.01}$ & 0.32$^{+0.28}_{-0.32}$ & - & -\\
NGC 5846 & 0.90$\pm{0.05}$ & 0.56$\pm{0.01}$ & 0.00$^{+0.09}_{-0.00}$ & - & - \\
         & $<$0.1 &  0.52$\pm{0.01}$ & 0.26$^{+0.13}_{-0.12}$ & 2.31$\pm{0.16}$ & (1.0) \\
NGC 6109 & 2.48$\pm{0.41}$ & 0.43$\pm{0.02}$ & 0.00$^{+0.16}_{-0.00}$ & - & - \\
         & 2.03$\pm{0.40}$ & 0.41$\pm{0.02}$ & 0.43$^{+0.15}_{-0.18}$ &
1.33$\pm{1.21}$ & (1.00) \\
NGC 6251 & $<$0.1 & 0.63$\pm{0.01}$ & 0.00$^{+0.04}_{-0.00}$ & - 
& - \\        
         & $<$0.1 & 0.39$\pm{0.01}$ & 0.13$^{+0.22}_{-0.13}$ & 0.23$\pm{0.01}$ & (1.0) \\ 
ARP 330  & $<$0.1 & 0.35$\pm{0.01}$ & 0.00$^{+0.20}_{-0.00}$ & - & - \\
NGC 6269 & 1.95$\pm{0.20}$ & 0.48$\pm{0.02}$ & 0.00$^{+0.17}_{-0.00}$ 
& - & - \\
         & 2.03$\pm{0.21}$ & 0.47$\pm{0.02}$ & 0.03$^{+0.15}_{-0.03}$ 
& $<$0.1 & (1.0) \\
NGC 6329 & 1.02$\pm{0.13}$ & 0.45$\pm{0.01}$ & 0.00$^{+0.15}_{-0.00}$ &
- & - \\
         & 1.04$\pm{0.13}$ & 0.44$\pm{0.01}$ & 0.00$^{+0.19}_{-0.00}$ &
$<$0.1 & (1.0) \\
NGC 6338 & $<$0.1 & 0.44$\pm{0.01}$ & 0.07$\pm{0.07}$ & - & - \\
         &  $<$0.1 & 0.50$\pm{0.01}$ & 0.00$^{+0.11}_{-0.00}$ & 0.38$\pm{0.04}$ & (1.0) \\
HCG 90   & 0.84$\pm{0.24}$ & 0.95$\pm{0.07}$ & 0.22$^{+0.27}_{-0.22}$ & 
- & - \\
UGC 12064& $<$0.1 & 0.38$\pm{0.03}$ & 0.04$^{+0.15}_{-0.04}$ & - & -
\\
         & $<$0.1 & 0.43$\pm{0.01}$ & 0.00$^{+0.16}_{-0.00}$ & 
$<$0.1 & (1.0) \\
HCG 92   & $<$0.1 & 0.43$\pm{0.03}$ & 0.35$^{+0.18}_{-0.24}$ & 
- & - \\
IC 1459  & 0.17$\pm{0.01}$ & 0.66$\pm{0.01}$ & 0.00$^{+0.04}_{-0.00}$ &
- & - \\
         & 0.15$\pm{0.02}$& 0.42$\pm{0.01}$ & 0.17$^{+0.19}_{-0.17}$ & 0.32$\pm{0.01}$ & (1.0) \\
NGC 7619 & 0.51$\pm{0.03}$ & 0.44$\pm{0.01}$ & 0.00$^{+0.09}_{-0.00}$ &
- & - \\
         & 0.57$\pm{0.03}$& 0.54$\pm{0.01}$ & 0.19$^{+0.07}_{-0.08}$& $<$0.1 
& (1.00) \\
HCG 97   & $<$0.1 & 0.41$\pm{0.01}$ & 0.00$^{+0.10}_{-0.00}$ & - & -
\\
         & $<$0.1 & 0.48$\pm{0.01}$ & 0.48$^{+0.16}_{-0.20}$ & 0.41$\pm{0.04}$ & 0.77$\pm{0.06}$  \\
NGC 7777 & $<$0.1 & 0.33$\pm{0.03}$ & (0.00) & - & - \\
\enddata
\end{deluxetable}

%% file: mulchaey.bbl
\begin{thebibliography}{}
\bibitem[Arnaud \& Evrard (1999)]{AE99} Arnaud, M., and Evrard, A. E. 1999, \mnras, 305, 631
\bibitem[Bauer \& Bregman (1996)]{BB96} Bauer, F., and Bregman, J. N. 1996,
\apj, 457, 382
\bibitem[Bauer et al. (2002)]{B02} Bauer, F. E., et al. 2002, \apj, in press
\bibitem[Bialek, Evrard \& Mohr (2001)]{BEM01} Bialek, J. J., Evrard, A. E., and
Mohr, J. J. 2001, \apj, 555, 597
\bibitem[Buote (1999)]{Buote99} Buote, D. A. 1999, \mnras, 309, 685
\bibitem[Buote (2000)]{Buote00} Buote, D. A. 2000, \mnras, 311, 176
\bibitem[Borgani et al. (2001)]{B01} Borgani, S. et al. 2001, \apj, 559, 71
\bibitem[Bryan (2000)]{bryan00} Bryan, G. L. 2000, \apj, 544, L1
\bibitem[Cavaliere, Menci \& Tozzi (1998)]{CMT98} Cavaliere, A., Menci, N., and 
Tozzi, P. 1998, \apj, 501, 493
\bibitem[Cowie, Henriksen \& Mushotzky (1987)]{c87} Cowie, L. L., Henriksen, M., and
Mushotzky, R. F. 1987, \apj, 317, 593
\bibitem[David et al. (1994)]{D94} David, L. P., Jones, C., Forman, W., and
Daines, S. 1994, \apj, 428, 544
\bibitem[Davis et al. (1995)]{D95} Davis, D. S., Mushotzky, R. F., Mulchaey,
 J. S., Worrall, D. M., Birkinshaw, M., and Burstein, D. 1995, \apj, 444, 582
\bibitem[Dickey \& Lockman (1990)]{DL90} Dickey, J. M., and Lockman, F. J. 1990,
\araa, 28, 215
\bibitem[Doe et al. (1995)]{Doe95} Doe, S. M., Ledlow, M. J., Burns, J. O., and
White, R. A. 1995, \aj, 110, 46
\bibitem[Dos Santos \& Mamon (1999)]{DM99} Dos Santos, S., and  Mamon, G. A. 
1999, \aap, 352, 1
\bibitem[Ebeling, Voges \& B\"{o}hringer (1994)]{E94} Ebeling, H., Voges, W.,
B\"{o}hringer, H. 1994, \apj, 436, 44
\bibitem[Evrard, Metzler \& Navarro (1996)]{ev96} Evrard, A. E., Metzler, C. A.,
Navarro, J. F. 1996, \apj, 469, 494
\bibitem[Fabricant \& Gorenstein (1983)]{fg83} Fabricant, D., and Gorenstein, P. 1983,
\apj, 267, 535
\bibitem[Fabricant et al. (1980)]{f80} Fabricant, D., Lecar, M., and Gorenstein, P. 1980,
\apj, 241, 552
\bibitem[Fabricant et al. (1984)]{f84} Fabricant, D., Rybicki, G., and Gorenstein, P. 1984,
\apj, 286, 186
\bibitem[Fang et al. (2002)]{F01} Fang, T., Marshall, H. L., Lee, J. C., Davis, D. S., and
Canizares, C. R. 2002, \apj, in press
\bibitem[Finoguenov, David \& Ponman (2000)]{FDP00} Finoguenov,A., David, L. P., 
Ponman, T. J. 2000, \apj, 544, 188
\bibitem[Fixsen et al. (1996)]{F96} Fixsen, D. J., Cheng, E. S., Gales, J. M.,
Mather, J. C., Shafer, R. A., Wright, E. L. 1996, \apj, 473, 576
\bibitem[Fujita \& Takahara (2000)]{FT00} Fujita, Y., and Takahara, F. 2000,
\apj, 536, 523
\bibitem[Fukugita, Hogan \& Peebles (1998)]{F98} Fukugita, M., Hogan, C. J., 
and Peebles, P. J. E. 1998, \apj, 503, 518
\bibitem[Garcia (1993)]{G93} Garcia, A. M. 1993, \aaps, 100, 47
\bibitem[Gehrels (1986)]{G86} Gehrels, N. 1986, \apj, 303, 336
\bibitem[Geller \& Huchra (1983)]{GH83} Geller, M. J., and Huchra, J. P. 1983,
\apjs, 52, 61
\bibitem[Hassinger et al. (1992)]{has92} Hasinger, G., Turner, T. J.,
George, I. M., and Boese, G. 1992, NASA/GSFC Office of Guest
Investigator Programs, Calibration Memo CAL/ROS92-001
\bibitem[Hellsten, Gnedin \& Miralda-Escude (1998)]{H98} Hellsten, U., Gnedin, N.Y.,
 and Miralda-Escude, J. 1998, \apj, 509, 56
\bibitem[Helsdon \& Ponman (2000a)]{HP00a} Helsdon, S. F., and Ponman, T. J. 
2000a,
\mnras, 315, 356
\bibitem[Helsdon \& Ponman (2000b)]{HP00b} Helsdon, S. F., and Ponman, T. J. 
2000b,
\mnras, 319, 933 
\bibitem[Henry et al. (1995)]{H95} Henry, J. P., Gioia, I. M., Huchra, J. P.,
Burg, R., McLean, B., et al. 1995, \apj, 449, 422
\bibitem[Hickson (1982)]{H82} Hickson, P. 1982, \apj, 255, 382
\bibitem[Hickson, Kindl \& Auman (1989)]{HKA89} Hickson, P., Huchra, J.,
and Kindl, E. 1989, \apjs, 70, 687
\bibitem[Hickson et al. (1992)]{H92} Hickson, P., Mendes de Oliveira, C., 
Huchra, J. P., and Palumbo, G. G. 1992, \apj, 399, 353
\bibitem[Horner (2001)]{H01} Horner, D. J. 2001, Ph D thesis, U. of Maryland
\bibitem[Huchra \& Geller (1982)]{HG82} Huchra, J. P., and Geller, M. J. 1982,
\apj, 257, 423
\bibitem[Hwang et al. (1999)]{Hw99} Hwang, U., Mushotzky, R. F., 
Burns, J. O., Fukazawa, Y, White, R. A. 1999, \apj, 516, 604
\bibitem[Kaastra \& Mewe (1993)]{KM93} Kaastra, J. S., and Mewe, R. 1993,
\aaps, 97, 443
\bibitem[Liedahl, Osterheld \& Goldstein (1995)]{L95} Liedahl, D. A.,
Osterheld, A. L., and Goldstein, W. H. 1995, \apj, 438, L 115
\bibitem[Lloyd-Davies, Ponman \& Cannon (2000)]{LD00} Lloyd-Davies, E. J.,
Ponman, T. J., and Cannon, D. B. 2000, \mnras, 315, 689
\bibitem[Loewenstein (2000)]{L00} Loewenstein, M. 2000, \apj, 532, 17
\bibitem[Maia, Da Costa \& Latham (1989)]{M89} Maia, M. A. G., da Costa, L. M., 
Latham, D. W. 1989, \apjs, 69, 809
\bibitem[Mahdavi et al. (1997)]{Ma97} Mahdavi, A., B\"{o}hringer, H., 
Geller, M. J., and Ramella, M. 1997, \apj, 483, 68
\bibitem[Mahdavi et al. (1999)]{Ma99} Mahdavi, A., Geller, M. J., B\"{o}hringer, H., 
Kurtz, M. J., and Ramella, M. 1999, \apj, 518, 69
\bibitem[Mahdavi et al. (2000)]{Ma00} Mahdavi, A., B\"{o}hringer, H.,
Geller, M. J., and Ramella, M. 2000, \apj, 534, 114
\bibitem[Mewe, Gronenschild \& van den Oord (1985)]{M85} Mewe, R.,
Gronenschild, E. H. B. M., and van den Oord, G. H. J. 1985, \aaps, 35, 503
\bibitem[Mohr, Mathiesen \& Evrard (1999)]{MME99} Mohr, J. J., Mathiesen, B., and Evrard, A. E. 1999,
\apj, 517, 627
\bibitem[Mulchaey (2000)]{Mul00} Mulchaey, J. S. 2000, \araa,
    38, 289
\bibitem[Mulchaey et al. (1993)]{M93} Mulchaey, J. S., Davis, D. S., 
Mushotzky, R. F., and Burstein, D. 1993, \apj, 404, L9
\bibitem[Mulchaey et al. (1996a)]{M96a} Mulchaey, J. S., Davis, D. S.,
Mushotzky, R. F., and Burstein, D. 1996a, \apj, 456, 80
\bibitem[Mulchaey et al. (1996b)]{M96b} Mulchaey, J. S., Mushotzky, R. F.,
Burstein, D., and Davis, D. S. 1996b, \apj, 456, L5
\bibitem[Mulchaey \& Zabludoff (1998)]{MZ98} Mulchaey, J. S., and 
Zabludoff, A. I. 1998, \apj, 496, 73
\bibitem[Navarro, Frenk \& White (1997)]{N97} Navarro, J. F., Frenk, C. S.,
and White, S. D. M. 1997, \apj, 490, 493
\bibitem[Nolthenius (1993)]{N93} Nolthenius, R. 1993, \apjs, 85, 1
\bibitem[Nolthenius \& White (1987)]{NW87} Nolthenius, R., and White, S. D. M. 
1987, \mnras, 225, 505
\bibitem[Perna \& Loeb (1998)]{PL98} Perna, R., and Loeb, A. 1998, \apj, 503, L135
\bibitem[Pierre, Bryan \& Gastaud (2000)]{PBG00} Pierre, M., Bryan, G., and 
Gastaud, R. 2000, \aap, 356, 403
\bibitem[Pildis, Bregman \& Evrard (1995)]{PBE95} Pildis, R. A., Bregman, J. N.,
and Evrard, A. E. 1995, \apj, 443, 514
\bibitem[Ponman \& Bertram (1993)]{PB93} Ponman, T. J., and Bertram 1993, 
Nature, 363, 51
\bibitem[Ponman et al. (1996)]{P96} Ponman, T. J., Bourner, P. D. J., 
Ebeling, H., B\"{o}hringer, H. 1996, \mnras, 283, 690
\bibitem[Ponman, Cannon \& Navarro (1999)]{P99} Ponman, T. J., Cannon, D. B., and
Navarro, J. F. 1999, Nature, 397, 135
\bibitem[Ribeiro et al. (1998)]{R98} Ribeiro, A. L., de Carvalho, R. R.,
Capelato, H. V., and Zepf, S. E. 1998, \apj, 497, 72
\bibitem[Saracco \& Ciliegi (1995)]{SC95} Saracco, P., and Ciliegi, P. 1995,
\aap, 301, 348
\bibitem[Savage et al. (2002)]{S02} Savage, B. D., Sembach, K. R., Tripp, T. M., and Richter, P. 2002,
\apj, 564, 631
\bibitem[Snowden et al. (1994)]{S94} Snowden, S. L., McCammon, D., Burrows, 
D. N.,
and Mendehall, J. A. 1994, \apj, 424, 714
\bibitem[Tully (1987)]{T87} Tully, R. B. 1987, \apj, 321, 280
\bibitem[Valageas \& Silk (1999)]{VS99} Valageas, P., and Silk, J. 1999, \aap,
350, 725
\bibitem[Werner, Worrall \& Birkinshaw (1999)]{WWB99} 
Werner, P. N., Worrall, D. M., and Birkinshaw, M. 1999, \mnras, 307, 722
\bibitem[Willmer et al. (1991)]{W91} Willmer, C. N. A., Focardi, P., Chan, R.,
Pellegrini, P. S., da Costa, N. L. 1991, \aj, 101, 57
\bibitem[Willmer et al. (1999)]{W99} Willmer, C. N. A., Maia, M. A. G.,
Mendes, S. O., Alonso, M. V., Rios, L. A., Chaves, O. L., and 
de Mello, D. F. 1999, \aj, 118, 1131
\bibitem[Yamada \& Fujita (2001)]{YM01} Yamada, M., and Fujita, Y. 2001,
\apj, 553, L145
\bibitem[Zabludoff \& Mulchaey (1998)]{ZM98} Zabludoff, A. I., and 
Mulchaey, J. S. 1998, \apj, 496, 39
\bibitem[Zabludoff \& Mulchaey (2000)]{ZM00} Zabludoff, A. I., and  
Mulchaey, J. S. 2000, \apj, 539, 136
\bibitem[Zimer, Zabludoff, \& Mulchaey (2002)]{ZZM02} Zimer, M., Zabludoff, A. I.,
 and Mulchaey, J. S. 2002, in preparation 
\end{thebibliography}
